\documentclass[a4paper,aps,superscriptaddress,floatfix,nofootinbib,twocolumn]{revtex4-1}

\usepackage{mathbbol}
\usepackage[pdftex]{graphicx}
\usepackage{latexsym,amsmath,verbatim,amssymb,txfonts}
\usepackage{color}
\usepackage{rotating}
\usepackage{verbatim}
\usepackage{multirow}
\usepackage[english]{babel}
\usepackage{comment}
\usepackage{hyperref}
\usepackage{dsfont}
\usepackage{bbold}
\usepackage{bbm}

\newcommand{\web}[1]{\href{https://coachscore.luddy.indiana.edu}{\nolinkurl{coachscore.luddy.indiana.edu}}~}

\begin{document}

%\title{Ranking sports coaches using networks: Who is the best?}
\title{Who is the best coach of all time? A network-based assessment of the career performance of professional sports coaches}

\author{\c{S}irag Erkol}
\affiliation{Center for Complex Networks and Systems Research, Luddy School
  of Informatics, Computing, and Engineering, Indiana University, Bloomington,
  Indiana 47408, USA}

\author{Filippo Radicchi}
\affiliation{Center for Complex Networks and Systems Research, Luddy School
  of Informatics, Computing, and Engineering, Indiana University, Bloomington,
  Indiana 47408, USA}
\email{filiradi@indiana.edu}

\begin{abstract}
We consider two large datasets consisting of all games played 
among top-tier 
%in major leagues of
European soccer 
clubs 
in the last $60$ years, and among professional American basketball teams in the past $70$ years. We leverage game data to build networks of pairwise interactions between the head coaches of the teams, and measure their career performance in terms of network centrality metrics. %The procedure allows us to establish rankings for specific leagues over customizable time windows. For instance, w
We identify \emph{Arsène Wenger}, \emph{Sir Alex Ferguson}, \emph{Jupp Heynckes}, \emph{Carlo Ancelotti}, and \emph{José Mourinho} as the top $5$ European soccer coaches of all time. In American basketball, the first $5$ positions of the all-time ranking are occupied by \emph{Red Auerbach}, \emph{Gregg Popovich}, \emph{Phil Jackson}, \emph{Don Nelson}, and \emph{Lenny Wilkens}. We further establish rankings by decade and season. We develop a simple methodology to monitor performance throughout a coach's career, and to dynamically compare the performance of two or more coaches at a given time. The manuscript is accompanied by the website \web  ~where complete results of our analysis are accessible to the interested readers.
\end{abstract}

\maketitle

%%%%%%%%%%%%%%%%%%%%%%%%%%%%%%%%%%%%%%%%%%%%%%%%%%%%%%%%%%%%%%%%%%%%%%%%%%%%%%%%%%%%%%%%%%%%%%%%%%%%%%%%%%%%%%%%%%%%%%%%%%%%%%%%%%%%%%%%%%%%%%%%%%%%%%%%%%%%%%%%%%%%%%%%%%%%%%%%%%%%%%%%%%%%%%%%%%%%%%%%%%%%%%%%%%%%%%%%%%%%%

\section{Introduction}

In a standard soccer league,
teams are ranked
on the basis of the total number of points they gather during the season.
Points are assigned to teams depending on the outcome of individual games
they take part in.
All teams play the same total number of games in the league by generally facing each other twice through the season. This is a simple, effective, and fair method to 
measure team performance in a single season. However, this standard metric does not differentiate the quality of a result depending on the opponent faced in the game. For example, beating a championship contender %team 
has the same importance as beating a team that %struggles and 
is positioned low in the league standings.
Also, the metric is not designed to properly measure performance over arbitrary time windows, e.g., a portion of a season or the aggregate of multiple seasons.

A simple way to partially address the above-mentioned issues and go beyond the mere counting of points is relying on a
macroscopic perspective of the league, where individual games are not seen as events that are independent one from the other, rather they are all seen as elementary contacts forming a complex network of interactions. The importance of a team in the web of contacts is self-established by the very structure of the network. The spirit is  similar to the one used in attempts of gauging the importance of web pages in information networks~\cite{brin1998anatomy, kleinberg1999authoritative}, establishing the relevance of papers~\cite{chen2007finding}, scientists~\cite{radicchi2009diffusion} and journals~\cite{bergstrom2008eigenfactor} in scientific networks, and measuring the influence of individuals~\cite{kempe2003maximizing, lu2016vital, erkol2019systematic, erkol2020influence} in social networks. Standard metrics of performance, such as the counting of points and/or wins, can be easily reconciled with local centrality metrics, e.g., in-degree and in-strength~\cite{barrat2004architecture, lu2016vital}. However, non-local metrics of centrality may allow to capture different aspects of performance~\cite{lu2016vital}. In particular, metrics such as the Bonacich~\cite{bonacich1987power} and the PageRank~\cite{brin1998anatomy} centralities allow to weigh the quality of wins and the quality of the opponents with simple, but reasonable self-consistent recipes.

Graph-based metrics of performance for teams and/or players have already been applied to
soccer~\cite{duch2010quantifying, buldu2019defining, PhysRevE.102.042120, grund2012network, sumpter2016soccermatics},~basketball \cite{skinner2010price, clauset2015safe, ribeiro2016advantage},~and various other sports \cite{park2005network,radicchi2011best, o2020complex, saavedra2010mutually, radicchi2012universality}.
The current paper explores the possibility of leveraging network centrality metrics to gauge career performance of coaches in two professional sports: soccer and basketball. There are no well-established metrics of performance for sports coaches, especially when the focus is on extended periods of time such as those corresponding to entire careers. 
%One possibility could be to simply count 
One could be tempted to evaluate career performance by simply counting
the number of trophies won by coaches. However, the importance of a trophy (e.g., an international cup) compared to another (e.g., a national championship) is hardly quantifiable, and, even for the same trophy, may vary from season to season depending on several factors. Also, the actual number of coaches with at least a trophy won during their career is a small fraction of the total number of coaches that managed professional teams, thus making the counting of trophies a recipe not very useful for the quantification of  performance for the vast majority of coaches. One could evaluate performance on other events, e.g., number of wins, rank positions in leagues, etc. However, the difficulty of properly quantifying the relative importance of the various elementary events would be exacerbated by a multitude of potential factors, e.g., type of competition, season, strength of the team trained, etc. The motivation behind our approach is indeed avoiding to make any complicated choice about the specific value to be assigned to the elementary events, and measure the performance of individual coaches in a self-contained manner by leveraging the structure of the head-to-head network among coaches.

We are not the first to consider the problem of measuring the performance of sports coaches. However,
the number of existing studies is quite limited. One of the most analyzed problems is the effect of sacking coaches, either during or at the end of a season, on the short- or long-term performance of teams~\cite{goff2019effect, madum2016managerial, besters2016effectiveness, muehlheusser2016impact}. %When it comes to analyzing career performances of coaches, % and comparing them with each other, 
%we are aware of two papers only. 
We are aware of only two papers  focusing on the analysis of coaches' career performance.
Xu \emph{et al.} use a Technique for Order of Preference by Similarity to Ideal Solution (TOPSIS) model to rank college basketball coaches~\cite{xu2015improved}. Hu \emph{et al.} consider data envelopment analysis and PageRank for ranking coaches in college sports~\cite{hu2015methods}.

As in Hu \emph{et al.}~\cite{hu2015methods}, also here we use PageRank as the main metric of performance for sports coaches. However, we differentiate from the work by Hu \emph{et al.} in two main respects. First, Hu \emph{et al.} arbitrarily weigh the importance of games on the basis of exogenous factors. In our approach, each game has the same \emph{a priori} importance; the effective value of a game is an emerging property of the system, depending on the quality of the opponents that are facing each other in the game.  Second, Hu \emph{et al.} consider datasets that regard games in the American college baseball, basketball and football leagues from 1990 to 2013. These data permit the construction of networks consisting of less than $100$ nodes. Our data span over temporal windows longer than $60$ years and allow us to build networks composed of more than $1,000$ nodes. The data further allows us to leverage dynamical ranking techniques to monitor entire career trajectories of many coaches.

%Summary of the paper
The paper is organized as follows. In section~\ref{methods}, we provide details of data selection, acquisition and curation, we describe how  information from the individual games is aggregated to form dynamic, directed, and weighted networks among coaches, and we illustrate the recipe at the basis of the network centrality metric used to measure the performance of coaches. In section~\ref{results}, we present our rankings of coaches. We consider all-time rankings, and top coaches of decades and seasons. Also, we use dynamic rankings to monitor career trajectories of individual coaches. In section~\ref{conclusions}, we provide our final considerations.
%, and indicate possible directions for future developments. 
Robustness of our results to some of the ingredients used in our ranking recipe are provided in appendix~\ref{appendix}. Additional results are available on the website \web~. The website allows interested readers to generate custom rankings, and to visualize career trajectories of all coaches included in our set of data.

%%%%%%%%%%%%%%%%%%%%%%%%%%%%%%%%%%%%%%%%%%%%%%%%%%%%%%%%%%%%%%%%%%%%%%%%%%%%%%%%%%%%%%%%%%%%%%%%%%%%%%%%%%%%%%%%%%%%%%%%%%%%%%%%%%%%%%%%%%%%%%%%%%%%%%%%%%%%%%%%%%%%%%%%%%%%%%%%%%%%%%%%%%%%%%%%%%%%%%%%%%%%%%%%%%%%%%%%%%%%%

\section{Methods}
\label{methods}

\subsection{Data}

Our analysis for comparing the performance of sports coaches is focused on two sports, men's soccer and basketball. 

\subsubsection{Soccer}

%In soccer, we will focus on the top tiers of top five leagues in European soccer: England, France, Germany, Italy, and Spain. 
%For soccer, 
Our main source of data is \url{transfermarkt.com}.
We collected publicly available information about the outcome of all games played by professional clubs in the top five
%\footnote{The ranking of the national leagues is based on the most recent country coefficients released by the UEFA.} 
national leagues of European soccer:  Premier League (England)~\cite{epl}, Ligue 1 (France)~\cite{ligue1}, Bundesliga (Germany)~\cite{bundesliga}, Serie A (Italy)~\cite{seriea}, and La Liga (Spain)~\cite{laliga}. The top five national leagues are selected on the basis of the most recent country coefficients released by UEFA~\cite{uefa}. We note that some of the league names changed during the period of time covered by our dataset. For example, the major English league was named ``First Division'' till season $1991/92$ and named ``Premier League'' since season $1992/93$. For each game, we collected information about the two teams playing the game, the outcome of the game, either a win by one of the two teams or a tie, and the day when the game was played. We uniquely identified all professional teams that took part in at least one edition of the above-mentioned leagues since season $1980/81$. For some leagues, we were able to trace back matches played since $1960/61$. Also, we included games of major European competitions that were played between the teams belonging to the top five European leagues. We considered the UEFA Champions League (previously named Champion Clubs' Cup), the UEFA Europa League (previously named UEFA Cup), the UEFA Cup Winners'  Cup, and the UEFA Super Cup (see  Table~\ref{table:abbreviations} for abbreviations).
For consistency with the data for the national leagues, we disregarded the results of extra times (or penalties),
%in direct-elimination matches
and only considered the results of the regular 90-minute time even if the games went to the extra times. The last season included in the dataset for all competitions is the $2019/20$ season. 

\begin{table}[!htb]
\begin{center}
\begin{tabular}{|l|l|r|r|r|r|}\hline 
%\begin{tabular}{|l|l|r|r|r|r|}\hline 
League & Country & Start %& End 
& Matches & Coaches \\\hline
Premier League & England & $1970/71$ %& 2019/20 
& $20, 713$ & $364$ \\\hline
Ligue 1 & France & $1980/81$  
%& 2019/20 
& $14, 709$ & $286$ \\\hline
Bundesliga & Germany & $1963/64$ 
%& 2019/20 
& $17, 375$ & $404$ \\\hline
Serie A & Italy & $1960/61$ 
%& 2019/20 
& $18, 144$ & $350$ \\\hline
La Liga &  Spain & $1960/61$ 
%& 2019/20 
& $20, 304$ & $492$ \\\hline
European cups & - & $1980/81$ 
%& 2019/20 
& $2, 043$ & $370$ \\\hline
\hline
Total & - & $1960/61$ %& 2019/20 
& $93, 288$ & $1, 777$ \\\hline
Combined & - & $1980/81$ %& 2019/20 
& $72, 981$ & $1,438$ \\\hline
\end{tabular}
\end{center}
\caption{
%{\bf Information on dataset content for soccer.} 
{\bf Summary table for the soccer dataset.} %We report basic information on the dataset considered in our study. 
From left to right, we report the name of the national league and the country of the league, the first season included in our dataset (all data have been collected till season $2019/20$), the number of matches of the league included in our data, and the number of coaches identified in the league. The dataset named ``Combined'' includes data from all the national leagues and European cups since season $1980/81$ only. This is the common period of coverage shared by all datasets of the individual national leagues and European competitions at our disposal. }
\label{table:1}
\end{table}
 
Please note that we did not include any games played for national cups nor games in minor European cups such as the UEFA Intertoto Cup. Also, no games played between national teams are included in our data. We stress that our selection excludes important national leagues, and glorious European soccer teams taking part in these leagues. The main reason behind our choice is the time coverage of the data from \url{transfermarkt.com}. For example, data for the Portuguese Primeira Liga~\cite{pri_liga} are reported only starting from season $1997/98$.

Still relying on data from \url{transfermarkt.com}, we determined the coaches that managed the teams playing each of the individual games of our sample. To this end, we gathered the coaching histories of all soccer teams in our dataset, and accounted for eventual changes of managers throughout the seasons.  We identified a few inconsistencies in the data from \url{transfermarkt.com}, that is, individual games of a team managed either by more than two coaches or no coach at all. We corrected those inconsistencies %manually
%Specifically, we gathered the coaching histories of all soccer teams in our sample. 
%, and merged these data with match results, which gave us head-to-head games between coaches. 
by relying on other sources of information, such as \url{bdfutbol.com} and \url{wikipedia.com}. After data curation, we were able to find unique coach-to-team assignments for more than $99\%$ of the games in our sample. 

\begin{figure}[!htb]
\centering
\includegraphics[width=.45\textwidth]{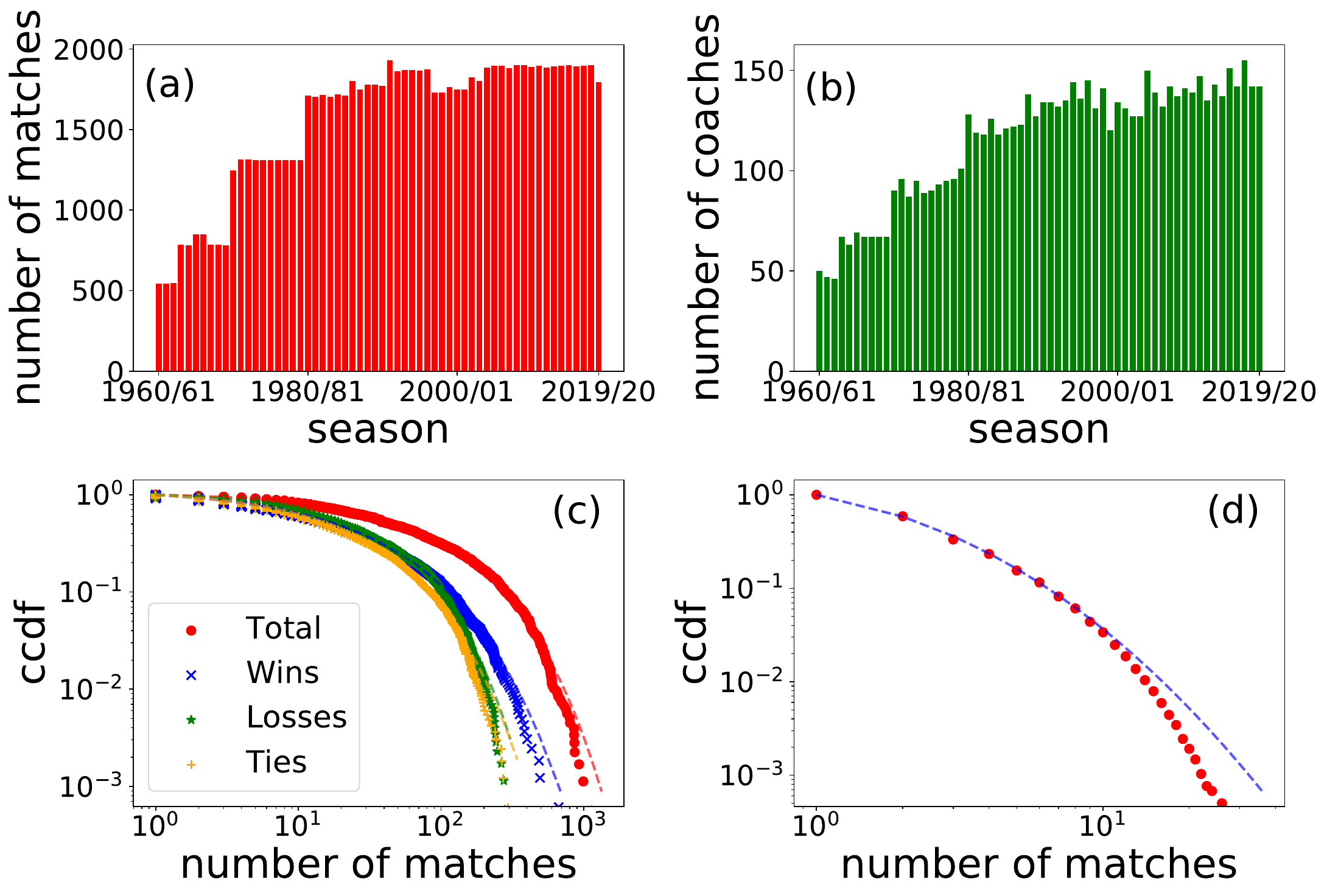}
\caption{{\bf Statistics of the soccer dataset.} (a) Total number of matches in our dataset per season. We combined together games from all the national leagues and European cups. (b) Number of coaches in our dataset per season. Each season includes every coach that has managed at least one game in that season. We combined the data of all games of the national leagues and European cups. (c) Complementary cumulative distribution functions (ccdf) of the number of matches managed, won, lost, and tied by coaches in our dataset. The stretched exponential ccdfs that best fit the empirical data are displayed as dashed curves. The stretched exponential ccdf is defined as $f(x)=\exp(-x^{\beta})$. The parameter of the best fits are $\beta=0.694$ for total matches, $\beta=0.591$ for wins, $\beta=0.788$ for losses, and $\beta=0.676$ for ties. (d) Ccdf for the number of head-to-head matches played between pairs of coaches. The best fit curve to the empirical data is also shown. In this case, we used the log-normal ccdf $g(x)=1 - \frac{1}{2} \, \textrm{erf} \left( \frac{\log (x) - \mu}{\sqrt{2}\, \sigma} \right)$. The parameters of the best fit are $\mu=0.453$ and $\sigma=0.944$.}
\label{fig:1}
\end{figure}

In total, we obtained information for $93, 288$ soccer matches. We identified $1,777$ unique  coaches who have managed at least one game. If we restrict our attention only to matches played from season $1980/81$ on, i.e., the period for which we have full information for all leagues considered in our study, then the total number of matches is $72,981$ and the total number of unique coaches is $1,438$ (see Table \ref{table:1} for details). As Figure~\ref{fig:1} shows, the number of games covered by our dataset consistently increases until $1980$, and stays more or less constant after that season. The same trend is observed for the number of coaches, although one could notice a slight increase even after $1980$. This fact indicates a growing tendency of  replacing coaches during the season.

We used the total number of games managed by a coach as a proxy for the career length of the coach (Figure~\ref{fig:1}). Empirical data are well described by a stretched exponential distribution\, whose parameter values are determined by maximum likelihood estimation~\cite{clauset2009power}. Stretched exponentials are good fits also for the distributions of total wins, losses and ties. Our finding is not compatible with previous results about career lengths of professional athletes, such as soccer, basketball, tennis, and baseball players~\cite{petersen2008distribution, petersen2011quantitative, radicchi2011best}, that are usually well fitted by power-law distributions. 
Our finding could reflect the fact that a coach's career is more resilient than an athlete's career. For instance, since there is only one coach in a team but many players, replacing a coach may be much more destabilizing for a team than replacing a player. Also, coaches have generally longer periods of apprenticeship than players. Only the very top coaches are able to become the managers of top-tier teams, thus leading to a selection bias towards a population with relatively homogeneous skills.
%\ch{The reason for the difference might be that coaches are more resilient to ``failures'' than athletes, as they might get more chances than athletes when they do not succeed in initial attempts. It might take longer to judge someone to be a bad coach than judge an athlete to be of low quality.} 

According to our dataset, Arsène Wenger %~\cite{wenger} 
tops the ranking in all four categories, with $1,339$ total games, $689$ wins, $307$ losses, and $343$ ties. We further measured the total number of head-to-head (h2h) games between pairs of coaches. Arsène Wenger and Sir Alex Ferguson top the ranking of h2h games with $36$ games played one against the other, followed by the pair Sir Alex Ferguson and Harry Redknapp with $33$ h2h games. Overall, empirical data are relatively well described by a log-normal distribution (parameters of the distribution are fitted using maximum likelihood estimation), but the tail of the empirical distribution is 
overestimated by the fitted log-normal distribution. (Figure~\ref{fig:1}).

\subsubsection{Basketball}

We collected results for the National Basketball Association (NBA)~\cite{nba} and the American Basketball Association (ABA)~\cite{aba} leagues from \url{basketball-reference.com}. NBA data start from season $1946/47$ (it was originally named as the Basketball Association of America, BAA) and end in season $2019/20$. We included regular-season and post-season games. Also, we included games of the ABA league, which co-existed with NBA between $1967/68$ and $1975/76$, until the two associations merged. We collected a total of $69, 549$ games (Table \ref{table:1_bball}). For each game, we collected information about the two teams playing one against the other, the outcome of the game, and the day when the game was played. As ties do not exist in basketball, we  consider the result after eventual overtimes as the actual game outcome. We further identified the coach of each team in the dataset, for a total of $364$ coaches.  

\begin{table}[!htb]
\begin{center}
\begin{tabular}{|l|r|r|r|r|}\hline 
%\begin{tabular}{|l|r|r|r|r|}\hline 
League & Start & End & Matches & Coaches \\\hline
NBA & $1946/47$ & $2019/20$ & $65, 400$ & $333$ \\\hline
ABA & $1967/68$ & $1975/76$ & $4, 149$ & $56$ \\\hline
\hline
Total & $1946/47$ & $2019/20$ & $69, 549$ & $364$ \\\hline
\end{tabular}
\end{center}
\caption{{\bf Summary table for the basketball dataset.}
%{\bf Information on dataset content for basketball.} 
%We report the information on the dataset used in the study. 
From left to right, we report the name of the league, the starting season, the ending season, the number of matches, and the number of coaches.}
\label{table:1_bball}
\end{table}

\begin{figure}[!htb]
\centering
\includegraphics[width=.45\textwidth]{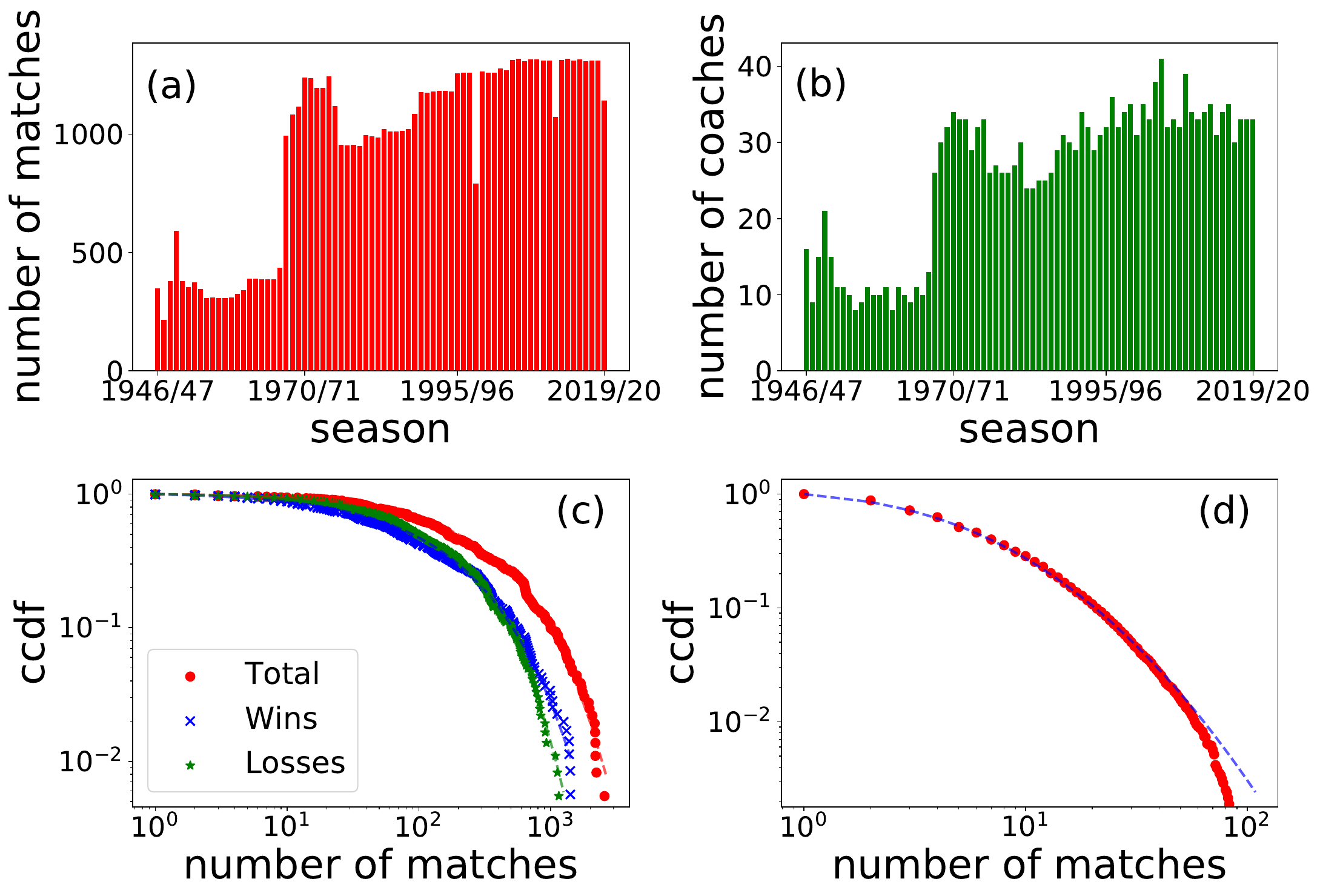}
\caption{{\bf Statistics of the basketball dataset.} (a) Total number of matches per season. (b) Number of coaches per season. Each season includes all coaches that have managed at least one game in that season. (c) Complementary cumulative distribution functions (ccdf) of the number of matches managed, won, and lost by coaches in our dataset. The stretched exponential ccdf is defined as $f(x)=\exp(-x^{\beta})$. The parameter of the best fits are $\beta=0.746$ for total matches, $\beta=0.664$ for wins, and $\beta=0.821$ for losses. (d) Ccdf for the number of head-to-head matches played between pairs of coaches. The best fit curve to the empirical data is also shown. In this case, we used the log-normal ccdf $g(x)=1 - \frac{1}{2} \, \textrm{erf} \left( \frac{\log (x) - \mu}{\sqrt{2}\, \sigma} \right)$. The parameters of the best fit are $\mu=1.567$ and $\sigma=1.098$.}
\label{fig:2}
\end{figure}

In Figure \ref{fig:2}, we see that both the number of games and coaches exhibit a sharp increase at the end of the $1960$s. This is when ABA started. After the merger of NBA and ABA in $1976$, the number of games and coaches decrease, then increase again, and finally reach a plateau. The plot of the number of games per season displays three clear dips. The $1998/99$ and $2011/12$ seasons were shorter due to NBA lockouts, and the $2019/20$ regular season was shorter due to the COVID-19 pandemic.

Also in Figure~\ref{fig:2}, we display the complementary cumulative distribution functions of the number of games, wins, and losses per coach, and we show the distribution of the number of h2h games between pairs of coaches. Data are well fitted by the same functions as those used in the analysis of soccer coaches (see Figure~\ref{fig:1}). Lenny Wilkens has the most matches coached and lost, respectively with $2,665$ games coached and $1,253$ losses. There is a tie between Larry Brown and Gregg Popovich for the highest number of wins, i.e., $1,447$ wins. In particular, Gregg Popovich has the highest number of  wins in NBA; however, considering the union of ABA and NBA games, Larry Brown equals Gregg Popovich in the number of wins.
The matchup that happened the most has been between Red Holzman and Gene Shue with $109$ h2h games, followed by Rick Adelman and Jerry Sloan with $104$ h2h games. 

\subsection{Networks of contacts among sports coaches}

We take advantage of the datasets described above to construct directed and weighted networks of contacts between coaches. A node in the network corresponds to a coach; pairwise interactions among coaches represent h2h games. In particular, networks are obtained by aggregating data about h2h games between pairs of coaches. A single data point is given by the game $g$ in which coach $i_g$ has a h2h game against $j_g$ at time $t_g$ (recall that time accuracy is one day). The contribution $c_{g\to ij}(t)$ at time $t \geq t_g$ of such a data point to the weight of the edge $ij$ is
\begin{equation}
    c_{g\to ij}(t) = \delta_{i_g,i} \, \delta_{j_g, j} \, e^{- \beta \, (t-t_g)} 
    %\ch{q}. 
    \, \times \, \left\{
    \begin{array}{ll}
         q_{\textrm{tie}} & \textrm{, if $i_g$ and $j_g$ tie in game $g$} \\
         q_{\textrm{loss}} & \textrm{, if $i_g$ loses against $j_g$ in game $g$}
    \end{array} \right.
    \; .
    \label{eq:ind_edge}
\end{equation}
Clearly, $c_{g\to ij}(t) = 0$ if $t < t_g$.
In Eq.~(\ref{eq:ind_edge}), the Kronecker $\delta$ function (i.e., $\delta_{x,y} = 1$ if $x=y$ and  $\delta_{x,y} = 0$, otherwise) tells us that a non-null contribution to the edge $ij$ requires that the game $g$ was indeed a h2h match between coaches $i$ and $j$.  The factor $e^{- \beta \, (t-t_g)}$ is an aging term identical to one used in Ref.~\cite{motegi2012network} for the dynamic ranking of tennis players. We consider two possible choices: $\beta=0$, meaning that the contribution of a game never ages; $\beta = 1/365$, meaning that the contribution is suppressed by a factor $e^{-1} \simeq 0.37$ every year. %\ch{$q=2$ if $i_g$ loses against $j_g$ in game $g$, and $q=1$ if $i_g$ and $j_g$ tie in the game.} 
The parameters $q_{\textrm{tie}}$ and $q_{\textrm{loss}}$ serve to weigh the contribution of the game outcome in the construction of the network.
We arbitrarily set $q_{\textrm{tie}} =1$ and $q_{\textrm{loss}} =2$ in most of the results of the paper.
Under %the choice of Eq.(\ref{eq:ind_edge}), 
this choice
a win counts twice as much as a tie (only for the directed edge from the loser to the winner), but a tie is counted twice (for the edges in both directions). In appendix~\ref{appendix}, we report results corresponding to the case $q_{\textrm{tie}} =1$ and $q_{\textrm{loss}} =3$ in which a win counts three times as much as a tie, but still a tie is counted twice. This choice is in line with how points are currently assigned in soccer leagues; however, the different choice does not significantly affect the outcome of our analysis (see Figure~\ref{fig:a2}). 

\begin{figure}[!htb]
\centering
\includegraphics[width=.45\textwidth]{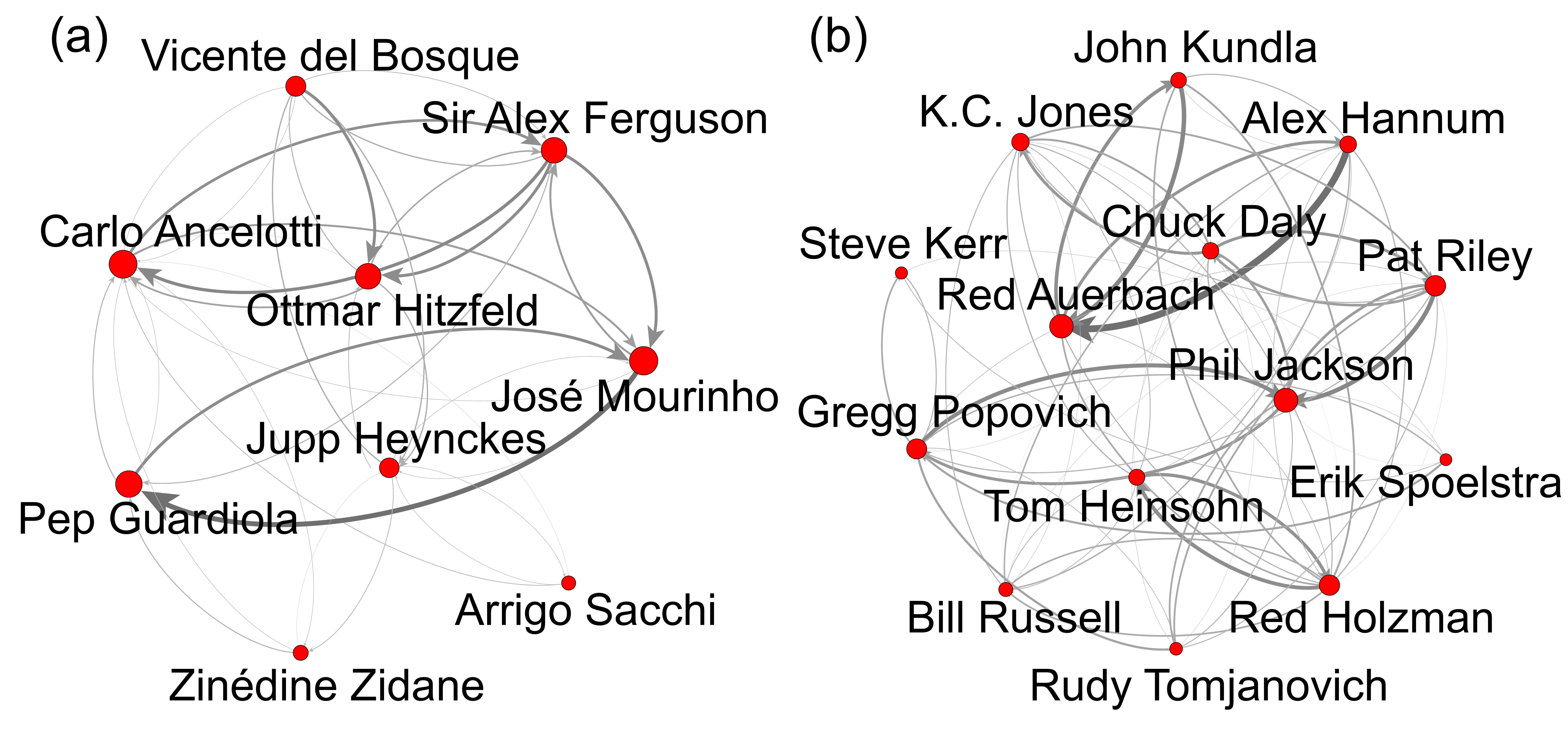}
\caption{{\bf Network of contact among top coaches.} (a) We display the network of contact among soccer coaches that have won the UEFA Champions League %or Champion Clubs' Cup 
at least twice after $1980$.
The network is built by setting $\beta =0$ in Eq.~(\ref{eq:ind_edge}) and including all games in our dataset.
In the visualization, the size of the nodes is proportional to their in-strength, and the width of the lines connecting pairs of nodes is proportional to the weight of the corresponding edge. (b) Same as in panel a, but for top basketball coaches. We visualize the network of contacts among coaches that have won at least $2$ NBA titles.}
\label{fig:nw_vis}
\end{figure}

The actual weight $w_{ij}(t)$ of the edge $ij$ at time $t$ is given by the sum of all individual contributions of the games in a given set $\mathcal{G}$, i.e.,
\begin{equation}
    w_{ij}(t) = \sum_{g \in \mathcal{G}} \, c_{g \to ij}(t) \; .
    \label{eq:weight}
\end{equation}
As apparent from Eq.~(\ref{eq:weight}), edge weights are dependent on the particular choice of the set of games $\mathcal{G}$ used in the construction of the network. As an example, in Figure~\ref{fig:nw_vis}, we display the networks of contacts restricted to subsets of top-tier soccer and basketball coaches. 
In what follows, we consider natural choices for such a set, as for example the set of all games played in specific national leagues and/or in specific seasons. 

\subsection{CoachScore}

Given a directed and weighted network composed of $N$ coaches and constructed according to the recipe of Eqs.~(\ref{eq:ind_edge}) and~(\ref{eq:weight}), we rank all coaches in the network using standard PageRank centrality~\cite{brin1998anatomy}. Given the context, we name the network centrality metric as CoachScore. Specifically, the CoachScore $p_i(t)$ of coach $i$ at time $t$ is computed as

\begin{equation}
\begin{aligned}
    p_i(t) =\alpha \, \sum_{j = 1}^N \frac{w_{ji}(t) \, p_j(t)}{ s^{\text{(out)}}_j(t)+\delta_{s_{j}^{\text{(out)}}(t), 0}}(1-\delta_{s_{j}^{\text{(out)}}(t), 0}) \\ + \frac{\alpha}{N} \, \sum_{j=1}^N \, p_j(t) \, \delta_{s_{j}^{\text{(out)}}(t), 0} + \frac{1-\alpha}{N}  \; .
    \label{eq:pagerank}
\end{aligned}
\end{equation}

The equation is valid for all nodes $i=1, \ldots, N$, with the constraint that $\sum_{i=1}^N p_i(t) = 1$.
$s^{\text{(out)}}_j(t) = \sum_{q=1}^N \, w_{jq}(t)$ is the so-called out-strength of coach $j$, i.e., the sum of the weights of all edges departing from node $j$~\cite{barrat2004architecture}.
%Eq.~(\ref{eq:pagerank}) assumes that no dangling nodes are present in the graph, i.e., $s^{(out)}_i(t) > 0$ for all $i$. This condition is always verified in the networks we consider in the paper, as no single coach has only wins in the datasets we analyze. 
The scores of the coaches are computed by iteration, starting from the suitable initial condition $p_i(t) = 1/N$ (although convergence of the algorithm does not require to start from such an initial condition).
%$0 \leq p_i(t) \leq 1$ for all $i$, and $\sum_{i=1}^N p_i(t) = 1$.  
Intuitively, each coach in the network carries a unit of ``prestige'' or ``credit,'' and we imagine 
that this quantity flows in the graph along its weighted connections. At each iteration of the algorithm, each coach $j$ distributes the entire credit to all its neighbors. The amount of credit given by coach $j$ to coach $i$ is proportional to the weight $w_{ji}(t)$. The term $\alpha \, \sum_{j = 1}^N \frac{w_{ji}(t) \, p_j(t)}{ s^{\text{(out)}}_j(t)+\delta_{s_{j}^{\text{(out)}}(t), 0}}(1-\delta_{s_{j}^{\text{(out)}}(t), 0})$ represents the portion of score received by coach $i$ from the immediate neighbors. 
In addition, each coach distributes part of the prestige equally to all the other coaches in the system, i.e., the term $\frac{1-\alpha}{N}$. Finally, the term $\frac{\alpha}{N}\sum_{j=1}^N \, p_j(t) \, \delta_{s_{j}^{\text{(out)}}(t), 0}$ on the rhs of Eq.~(\ref{eq:pagerank}) serves as a correction for the case of dandling nodes, i.e.,  nodes with null out-strength, which otherwise would behave as sinks in the diffusion process. The system of Eq.~(\ref{eq:pagerank}) converges (within a priori fixed precision $\epsilon$, here we set $\epsilon = 10^{-6}$) after a certain number of iterations of the algorithm. The resulting score $p_i(t)$ quantifies the relative credit that coach $i$ has at time $t$. The factor $0 \leq \alpha \leq 1$ determines the relative importance between local diffusion of prestige among immediate neighbors, and global redistribution of credit to the entire network. In our calculations, we choose the customary value $\alpha = 0.85$. 

If the network is constructed by setting $\beta=0$ in Eq.~(\ref{eq:ind_edge}), %the score of each coach is a measure of performance for their career up to the time $t$ when the score is measured. For $\beta=0$ in fact, 
all games in the input set $\mathcal{G}$ (up to the time $t$ when the score is actually measured)
%in the career of the coach 
have the same weight in the determination of the score of a coach. 
Instead for $\beta=1/365$, the score of a coach is mainly determined by the games close to the time $t$ when the score is quantified.

CoachScore is highly correlated with the number of wins (see Figure~\ref{fig:a1}).
High correlation between CoachScore and local centrality metrics, e.g., the in-strength (sum of the weights of the incoming connections of a node) is apparent too~\cite{radicchi2011best}.  With respect to local centrality metrics, however, CoachScore has the advantage of giving high importance to wins (and ties) against quality opponents, where quality is self-consistently quantified by CoachScore. We do not claim that CoachScore is a better metric of performance than other metrics. It just provides a way of measuring the performance of sports coaches different from the simple enumeration of wins.

%%%%%%%%%%%%%%%%%%%%%%%%%%%%%%%%%%%%%%%%%%%%%%%%%%%%%%%%%%%%%%%%%%%%%%%%%%%%%%%%%%%%%%%%%%%%%%%%%%%%%%%%%%%%%%%%%%%%%%%%%%%%%%%%%%%%%%%%%%%%%%%%%%%%%%%%%%%%%%%%%%%%%%%%%%%%%%%%%%%%%%%%%%%%%%%%%%%%%%%%%%%%%%%%%%%%%%%%%%%%%

\section{Results}
\label{results}

We present results obtained by ranking coaches on the basis of their CoachScore values. The 
set $\mathcal{G}$ of games used to construct the weighted network that serves for the computation of the centrality metric is the main ingredient we play with. Most of our results are obtained by setting $\beta = 0$ in Eq.~(\ref{eq:ind_edge}) while aggregating games to build the network. Also, we consider the setting $\beta = 1/365$ 
as a simple way to define a dynamical score useful to monitor the career evolution of coaches over time. 
In the following, we present results first for soccer coaches, and then for basketball coaches.

\subsection{Soccer}

\subsubsection{Top coaches of all time}

First, we report on the all-time rankings at the national level. We consider all games of the national leagues listed in Table~\ref{table:1}, and construct weighted networks using games of national leagues only as the set $\mathcal{G}$ in Eq.~(\ref{eq:weight}). In particular  while computing edge weights, we set $\beta=0$ and $t$ equal to the day of the most recent game in $\mathcal{G}$ in Eq.~(\ref{eq:ind_edge}).

\begin{table}[!htb]
\begin{center}
\resizebox{.48\textwidth}{!}{%
\begin{tabular}{|r|l|l|l|l|l|l|}\hline 
%\begin{tabular}{|r|l|l|l|l|l|l|}\hline 
Rank & England & France & Germany & Italy & Spain & Combined \\\hline
\multirow{2}{*}{1} & Sir Alex & Guy & Otto & Giovanni & Luis & Arsène \\
& Ferguson & Roux & Rehhagel & Trapattoni & Aragonés & Wenger \\\hline
\multirow{2}{*}{2} & Arsène & Claude & Jupp & Carlo & Miguel & Sir Alex \\
& Wenger & Puel & Heynckes & Mazzone & Munoz & Ferguson \\\hline
\multirow{2}{*}{3} & Brian & Jean & Udo & Nils & Javier & Jupp \\
& Clough & Fernandez & Lattek & Liedholm & Irureta & Heynckes \\\hline
\multirow{2}{*}{4} & Sir Bobby & Joël & Thomas & Carlo & Ernesto & Carlo \\
& Robson & Muller & Schaaf & Ancelotti & Valverde & Ancelotti \\\hline
\multirow{2}{*}{5} & Harry & Frédéric & Ottmar & Fabio & Víctor & José \\
& Redknapp & Antonetti & Hitzfeld & Capello & Fernandez & Mourinho \\\hline
\multirow{2}{*}{6} & José & Rolland & Felix & Nereo & Diego & Claudio \\
& Mourinho & Courbis & Magath & Rocco & Simeone & Ranieri \\\hline
\multirow{2}{*}{7} & David & Jean-Claude & Erich & Luciano & Joaquín & Otto \\
& Moyes & Suaudeau & Ribbeck & Spalletti & Caparrós & Rehhagel \\\hline
\multirow{2}{*}{8} & Ron & Élie & Hennes & Francesco & John & Pep \\
& Atkinson & Baup & Weisweiler & Guidolin & Toshack & Guardiola \\\hline
\multirow{2}{*}{9} & Sam & Rudi & Christoph & Luigi & Javier & Jürgen \\
& Allardyce & Garcia & Daum & Radice & Clemente & Klopp \\\hline
\multirow{2}{*}{10} & Sir Kenny & Jacques & Dieter & Helenio & Gregorio & Guy \\
& Dalglish & Santini & Hecking & Herrera & Manzano & Roux \\\hline
\end{tabular}
}
\end{center}
\caption{{\bf Top $10$ coaches in European soccer of all time.} We report the $10$ best coaches for each of the national leagues we consider in this paper (see Table~\ref{table:1}). In the rightmost column, we report the top $10$ coaches obtained on the basis of the combination of all games, national and international, at our disposal since season $1980/81$.}
\label{table:2}
\end{table}

The top $10$ all-time coaches of each national league are reported in Table~\ref{table:1}. In the English Premier League, Sir Alex Ferguson, the legendary coach of Manchester United FC for more than $25$ years and winner of $13$ national championships, is at the top of the ranking. For the French Ligue 1, Guy Roux is ranked number $1$. He was the coach of AJ Auxerre for roughly $40$ years, winning $1$ championship in $1995/96$. Otto Rehhagel, winner of $3$ league titles, sits in the first place of the German Bundesliga. Giovanni Trapattoni tops the ranking of the Italian Serie A. ``Il Trap'' won $7$ Italian championships. Finally, in the Spanish La Liga, Luis Aragon\'es turns out to be the best performing coach. He was the manager of several teams in Spain, and won $1$ championship with Atl\'etico Madrid in $1976/77$.

We note that career length is a quite important factor for PageRank. 
This type of dependence of the PageRank metric in dynamic/growing networks is well known~\cite{medo15, medo17}. Also in our case, the age dependence of the centrality metric is a natural consequence of the fact that all games are aggregated together in a memory-less fashion, and coaches that managed teams for tens of seasons are represented by nodes that are very well connected, thus highly central, in the graph of contacts. Carlo Mazzone, Luciano Spalletti, and Francesco Guidolin, for example, all managed Italian Serie A teams for $20$ years or more, but they never won a national championship in Italy. We believe that performance is a multidimensional metric, and career length should be seen as one of its dimensions. Indeed, the ability of a coach to remain active for many years is certainly an uncommon skill (see Figure~\ref{fig:1}). We stress that the use of the PageRank metric based on the aggregation of multiple decades of data clearly penalizes coaches that had short, even if successful, careers. Coaches that are not well represented by our datasets, for example because they are still coaching today or they managed teams earlier than the starting date of the matches covered by our datasets, are penalized too.

Also, we establish the ranking of soccer coaches based on their overall careers by considering the so-called ``combined'' dataset, which consists of all national and international games since season $1980/81$ (see Table~\ref{table:1}). Maybe surprisingly, Arsène Wenger comes at the top of this ranking. He has been the coach of AS Nancy and AS Monaco FC. More notably, Wenger has been the manager of Arsenal FC for more than $20$ years, achieving many successes.
He is followed by Sir Alex Ferguson. Jupp Heynckes occupies the third place. Heynckes has coached important clubs such as Real Madrid CF and FC Bayern Munich, winning $4$ national league and $2$ Champions League titles. As for the overall national rankings, still here career length is extremely valued. However, the international nature of the careers is valued too. We see that the top $10$ rank positions are occupied by several coaches who successfully managed teams in different European countries. A paradigmatic example is Claudio Ranieri, who started his career in the late $1980$s and coached teams in all major leagues considered in this study except for the German Bundesliga. Ranieri won the Premier League title in $2015/16$ with Leicester City FC. Finally, a special mention is necessary for José Mourinho. Our results are based on a dataset that does not include any of his games for seasons $2002/03$ and $2003/04$, when he was the head coach of FC Porto. We stress that he won the national championships in both seasons; he further won the UEFA Cup in $2002/03$ and the UEFA Champions League in $2003/04$. Given the importance of the coach for European soccer, we performed a separate analysis by including in the combined dataset all games of the Portuguese Primeira Liga~\cite{pri_liga} and games of the European competitions involving Portuguese teams for the two aforementioned seasons. The actual value of the score for José Mourinho increases by more than $15\%$. However, the increment is not sufficient to let him gain any position in the all-time ranking.

\subsubsection{Top coaches of the decade}

We repeated a similar analysis by dividing game data in decades. These sets of games are subsamples of the sets considered in the section above when establishing the all-time rankings. For example, the $1960$s decade of Serie A consists of all games played in the Italian Serie A in the $10$ consecutive seasons ranging from $1960/61$ to $1969/70$. We could not consider some combinations of league/decade for lack of data, e.g., Ligue 1 in the $1960$s. For the Bundesliga in the $1960$s, the ranking is established only on the seven seasons at our disposal. We remark that the specific choice made here for the selection of the games that contribute to the creation of the networks still favors some coaches with respect to others. For example, the $1970$s performance of a coach with career spanning $10$ consecutive seasons from $1965/66$ till $1974/75$ is unavoidably penalized compared to the performance of a coach with a career of identical length but spanning from season $1970/71$ to season $1979/80$.

\begin{table}[!htb]
\begin{center}
\resizebox{.48\textwidth}{!}{%
\begin{tabular}{|r|l|l|l|l|l|l|}\hline 
%\begin{tabular}{|r|l|l|l|l|l|l|}\hline 
Decade & Premier League & Ligue 1 & Bundesliga & Serie A & La Liga & Combined \\\hline
\multirow{2}{*}{1960s} & - & - & Helmuth & Helenio & Miguel & - \\
&  &  & Johannsen & Herrera & Munoz &  \\\hline
\multirow{2}{*}{1970s} & Dave & - & Udo & Nils & Carriega & - \\
& Sexton &  & Lattek & Liedholm &  &  \\\hline
\multirow{2}{*}{1980s} & Brian & Aimé & Jupp & Giovanni & Javier & Jupp \\
& Clough & Jacquet & Heynckes & Trapattoni & Clemente & Heynckes \\\hline
\multirow{2}{*}{1990s} & Sir Alex & Guy & Otto & Marcello & Javier & Sir Alex \\
& Ferguson & Roux & Rehhagel & Lippi & Irureta & Ferguson \\\hline
\multirow{2}{*}{2000s} & Sir Alex & Claude & Thomas & Carlo & Joaquín & Sir Alex \\
& Ferguson & Puel & Schaaf & Ancelotti & Caparrós & Ferguson \\\hline
\multirow{2}{*}{2010s} & Arsène & Christophe & Dieter & Massimiliano & Diego & Pep \\
& Wenger & Galtier & Hecking & Allegri & Simeone & Guardiola \\\hline
\end{tabular}
}
\end{center}
\caption{{\bf Top European soccer coaches by decade.} For each decade, we report the best coach of each of the national leagues we consider in this paper (see Table~\ref{table:1}). Empty cells indicate that no data are at our disposal for the corresponding combination of league/decade. In the rightmost column, we report the best coach of each decade obtained on the basis of the combination of all data at our disposal, including national and international competitions.}
\label{table:3}
\end{table}

With those considerations in mind, we first constructed networks by setting the parameter $\beta =0$, and then evaluated the CoachScore of each coach in the network. We remark that the network is constructed at a time $t$ equal to the day of the most recent game in the set, so that all games of the input set $\mathcal{G}$ are aggregated together to form the corresponding network. Top coaches by decade are reported in Table~\ref{table:3}. Several of the coaches already present in the all-time top $10$ ranking appear here too. 
%We do not provide details about coaches placed in the top $10$ national rankings here. 
At the continental level, Jupp Heynckes tops the $1980$s ranking, Sir Alex Ferguson is elected as the best coach of the $1990$s and $2000$s, and Pep Guardiola is identified as the best coach of the past decade.

\subsubsection{Top coaches of the season}

\begin{figure}[!htb]
\centering
\includegraphics[width=.45\textwidth]{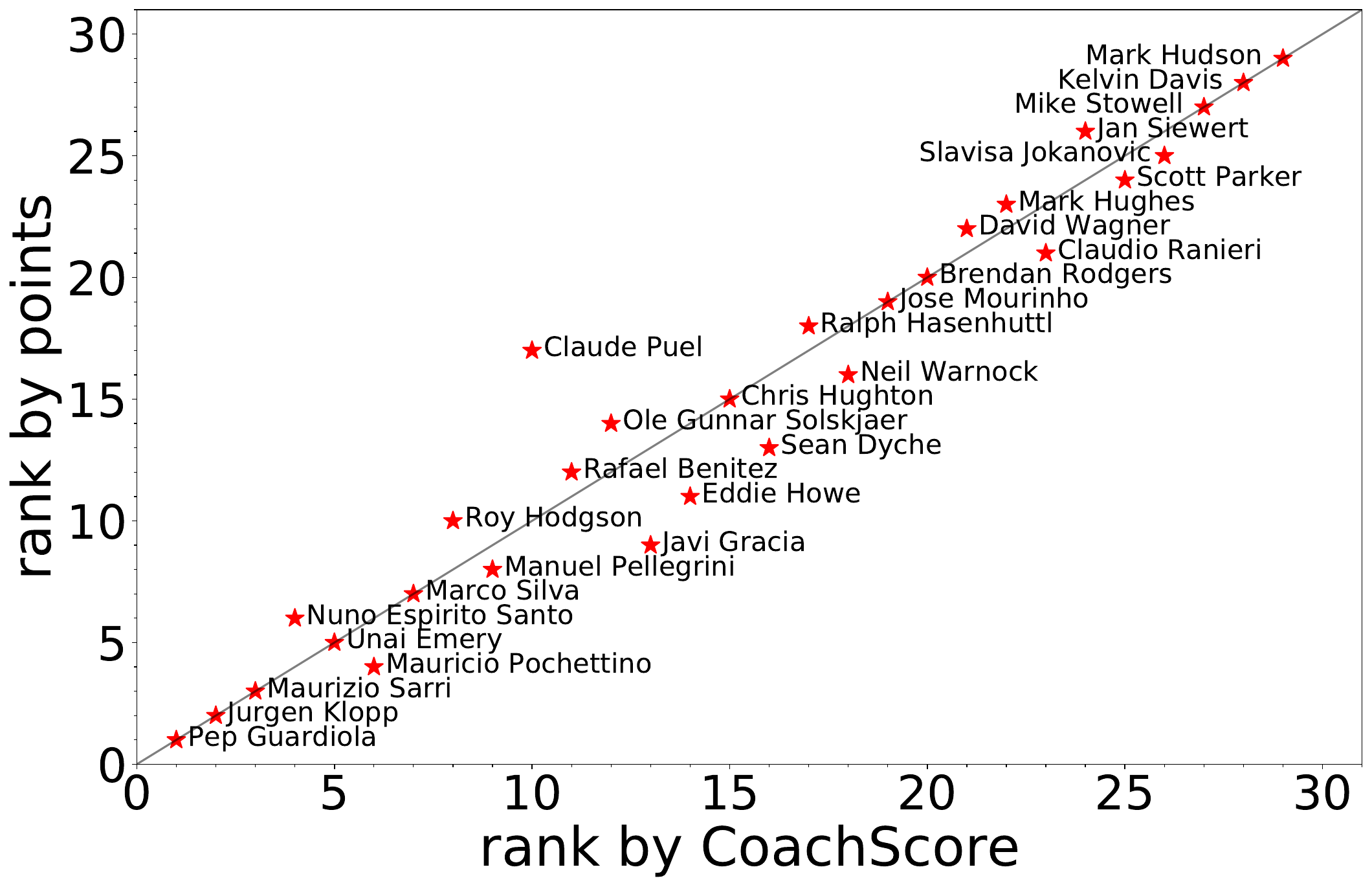}
\caption{{\bf CoachScore {\it vs.} number of points.} We report the rank positions for soccer coaches who managed teams in the $2018/19$ English Premier League. We rank coaches on the basis of their CoachScore and compare those ranks with those obtained with the number of points they gathered in the league. Spearman rank correlation coefficient is $\rho=0.973$, and Kendall rank correlation coefficient is $\tau=0.896$. The line stands for perfect agreement between the two rankings. Note that some of the coaches did not manage their team for the entire season. For example, both Jos\'e Mourinho and Ole Gunnar Solskj{\ae}r appear in the plot as they both served as head coaches for Manchester United FC during  the $2018/19$ season.}
\label{fig:4}
\end{figure}

Finally, we establish rankings for individual seasons. Networks are built by selecting games played in a given season only. Weights of the network edges are still calculated by setting the parameter $\beta = 0$ and using the $t$ value of the most recent game in the set. In Tables \ref{table:4}-\ref{table:4a}, we list the top coaches for all seasons covered by our dataset. We see many of the coaches already listed in the top rankings of all time and by decade. Some of them are elected as the best coach for multiple seasons.  In general, we note that the coach of the team winning the national league tops the CoachScore ranking of the season too. Points made in the league and CoachScore are indeed highly correlated, as for instance shown in Figure~\ref{fig:4} for the $2018/19$ English Premier League. There are, however, several seasons where the elected best coach according to CoachScore is not the one who actually won the national championship. Examples from the Italian Serie A are Giancarlo De Sisti in $1981/82$, Sven-Göran Eriksson in $1985/86$, Fabio Capello in $2001/02$, and Claudio Ranieri in $2009/10$, whose teams ranked second in the league only one point behind the season champions. Similar ``anomalies'' are present in the other national leagues too. In the rankings based on the combination of national and European games of a season, we see that the top coach is generally the Champions League winner (in some cases, the winner of the Europa League too). Also, here there are some anomalies, in the sense that the top European coach of the season won only the national championship and reached the final stages of a European competition without winning it. Examples are Diego Simeone in $2013/14$, Luis Enrique in $2015/16$, and Ernesto Valverde in $2018/19$. There are also cases where the top European coach neither won the national league nor a European competition, such as Javier Irureta in $2001/02$, and Julian Nagelsmann in $2019/20$.

\subsubsection{Dynamic ranking}

We now turn our attention on the dynamical version of the ranking by building weighted networks where we account for the aging of the contribution of individual games. To this end, we set $\beta = 1/365$ in Eq.~(\ref{eq:ind_edge}). At each point in time $t$, we reconstruct the network, and recompute the CoachScore rank of all coaches in the network.

We use dynamic ranking to first establish the best coaches at time of writing of this paper. Please note that no games of season $2020/21$ are included, thus our current ranking is based on games played till August $2020$. In Table~\ref{table:5}, we list the top $10$ coaches at the national and continental levels. Jürgen Klopp is the best coach in the English Premier League and in Europe. Christophe Galtier, Julian Nagelsmann, Gian Piero Gasperini, and Diego Simeone top their national rankings. 

\begin{table}[!htb]
\begin{center}
\resizebox{.48\textwidth}{!}{%
\begin{tabular}{|r|l|l|l|l|l|l|}\hline 
%\begin{tabular}{|r|l|l|l|l|l|l|}\hline 
Rank  & Premier League & Ligue 1 & Bundesliga & Serie A & La Liga & Combined \\\hline
\multirow{2}{*}{1} & Jürgen & Christophe & Julian & Gian Piero & Diego & Jürgen \\ 
 & Klopp & Galtier & Nagelsmann & Gasperini & Simeone & Klopp \\\hline
\multirow{2}{*}{2} & Pep & Thomas & Lucien & Simone & Zinédine & Pep \\ 
 & Guardiola & Tuchel & Favre & Inzaghi & Zidane & Guardiola \\\hline
\multirow{2}{*}{3} & Ole Gunnar & David & Peter & Maurizio & Ernesto & Diego \\ 
 & Solskjaer & Guion & Bosz & Sarri & Valverde & Simeone \\\hline
\multirow{2}{*}{4} & José & Michel Der & Christian & Stefano & José Luis & Julian \\ 
 & Mourinho & Zakarian & Streich & Pioli & Mendilibar & Nagelsmann \\\hline
\multirow{2}{*}{5} & Sean & Rudi & Hans-Dieter & Antonio & Quique & Thomas \\ 
 & Dyche & Garcia & Flick & Conte & Setién & Tuchel \\\hline
\multirow{2}{*}{6} & Nuno Espírito & Stéphane & Marco & Sinisa & Pepe & Lucien \\ 
 & Santo & Moulin & Rose & Mihajlovic & Bordalás & Favre \\\hline
\multirow{2}{*}{7} & Roy & Leonardo & Adi & Roberto De & Javier & Maurizio \\ 
 & Hodgson & Jardim & Hütter & Zerbi & Calleja & Sarri \\\hline
\multirow{2}{*}{8} & Ralph & Thierry & Dieter & Gennaro & Paco & José \\ 
 & Hasenhüttl & Laurey & Hecking & Gattuso & López & Mourinho \\\hline
\multirow{2}{*}{9} & Frank & Julien & Florian & Massimiliano & Imanol & Carlo \\ 
 & Lampard & Stéphan & Kohfeldt & Allegri & Alguacil & Ancelotti \\\hline
\multirow{2}{*}{10} & Mauricio & Patrick & Niko & Walter & Gaizka & Zinédine \\ 
 & Pochettino & Vieira & Kovac & Mazzarri & Garitano & Zidane \\\hline
\end{tabular}
}
\end{center}
\caption{{\bf Current list of top coaches in European soccer.} We report the $10$ currently best coaches for each of the national leagues we consider in this paper. In the rightmost column, we report the top $10$ coaches obtained on the basis of the combination of all national and international games at our disposal. Weights of network edges are computed at the end of season $2019/20$. We use $\beta = 1/365$ in Eq.~(\ref{eq:ind_edge}).}
\label{table:5}
\end{table}

\begin{figure}[!htb]
\centering
\includegraphics[width=.45\textwidth]{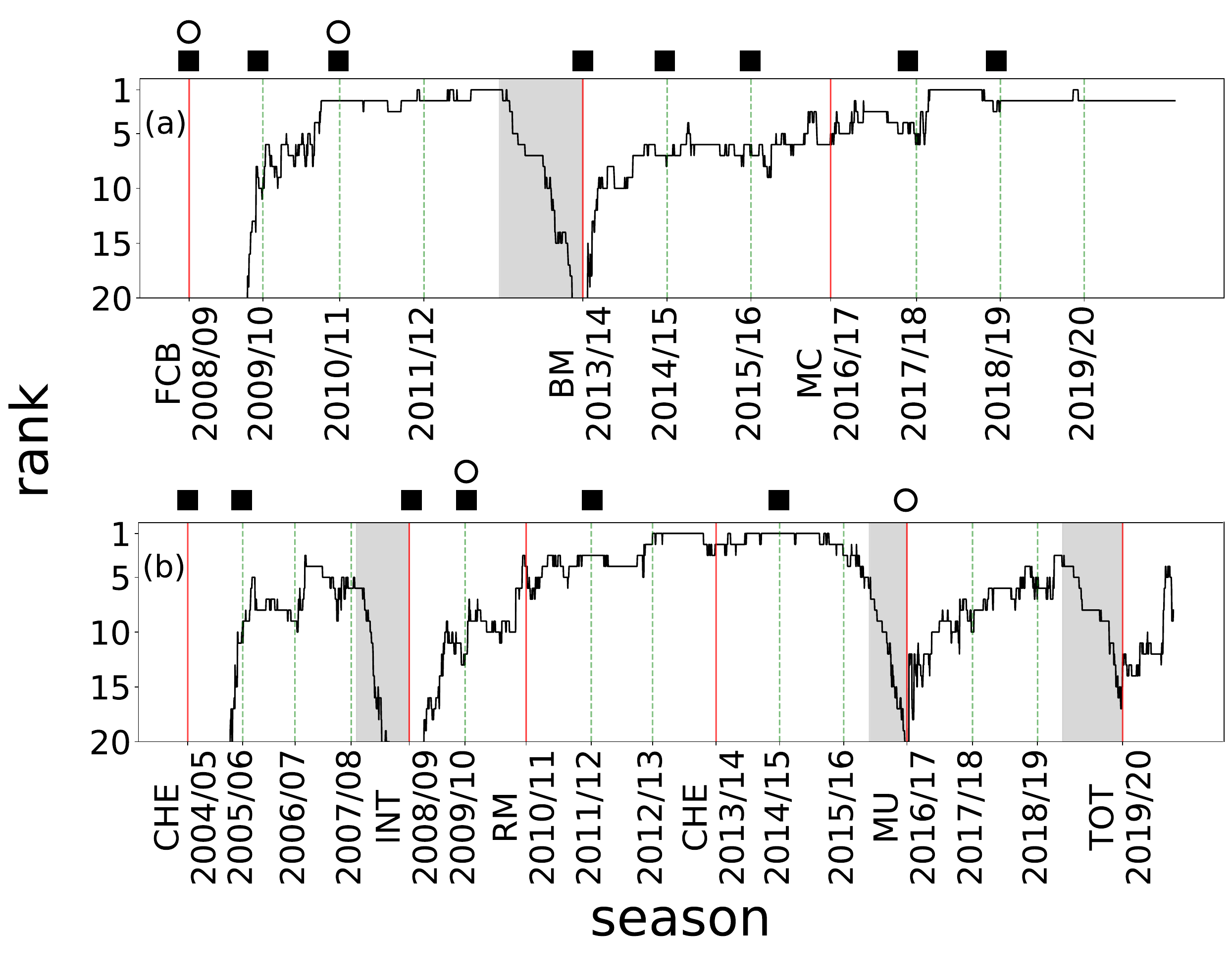}
\caption
{{\bf Monitoring the career performance of soccer coaches.} (a)
We visualize the dynamic rank of Pep Guardiola throughout his career. Rank positions are updated after each game day. Red lines indicate when the coach takes over a new team, dashed green lines indicate the start of a new season with the current team, and grey shaded areas represent periods of time when the coach is not managing any team. For each season of the coach's career, we draw a full black square to indicate the victory of the national championship, and an empty circle to represent the win of a European cup. FCB stands for FC Barcelona, BM for FC Bayern Munich, and MC for Manchester City FC. (b) Same as in panel a, but for José Mourinho. CHE stands for Chelsea FC, INT for FC Internazionale Milano, RM for Real Madrid CF, MU for Manchester United FC, and TOT for Tottenham Hotspur FC.}
\label{fig:5}
\end{figure}

Also, we take advantage of dynamic CoachScore to monitor performances of coaches throughout their careers. We use the combination of all national and continental games to construct networks. In Figure~\ref{fig:5}, we display the career trajectories of Pep Guardiola and José Mourinho. 

Guardiola started his career at FC Barcelona in $2008/09$. According to our metric, he  enters in the top $10$ ranking at the end of his first season, in the top $5$ list at the end of the second season, and tops the ranking during the $2011/12$ season. During the ``sabbatical'' $2012/13$ season, he loses rank positions. From $2013/14$ till $2015/16$ he was the coach of FC Bayern Munich. In spite of being consistently ranked in the top $10$, he never reaches the actual top of the ranking during that period. Finally, he started coaching Manchester City FC in $2016/17$. His dominant performance in the English Premier League makes him consistently ranked in the top $5$ coaches in Europe. He is ranked at the top position for a great part of season $2017/18$, and currently sits at position number $2$ right behind Jürgen Klopp. 

The career trajectory of José Mourinho does not include any of the seasons prior to $2004/05$, when he started coaching Chelsea FC. We remind the reader that he was the head coach of FC Porto in seasons $2002/03$ and $2003/04$. He won the national championships in both seasons; he further won the UEFA Cup in $2002/03$ and the UEFA Champions League in $2003/04$. The very fact that these data points are not included in our dataset clearly penalizes his performance as measured in the all-time and $2000$s rankings. Also, it affects the effective performance measured by the dynamic CoachScore at the beginning of his tenure as the head coach of Chelsea FC. We see, however, that thanks to his excellent performance at Chelsea FC he is ranked in the top $10$ between $2005$ and $2007$. While coaching FC Internazionale Milano, Real Madrid CF, and, for the second time, Chelsea FC, he is steadily ranked in the top $10$. He is ranked in the top $10$ also during his recent tenures at Manchester United FC and Tottenham Hotspur FC. Clear drops in rank positions are visible only during the three breaks he had in part of the seasons $2007/08$, $2015/16$ and $2018/19$.

\begin{figure}[!htb]
\centering
\includegraphics[width=.45\textwidth]{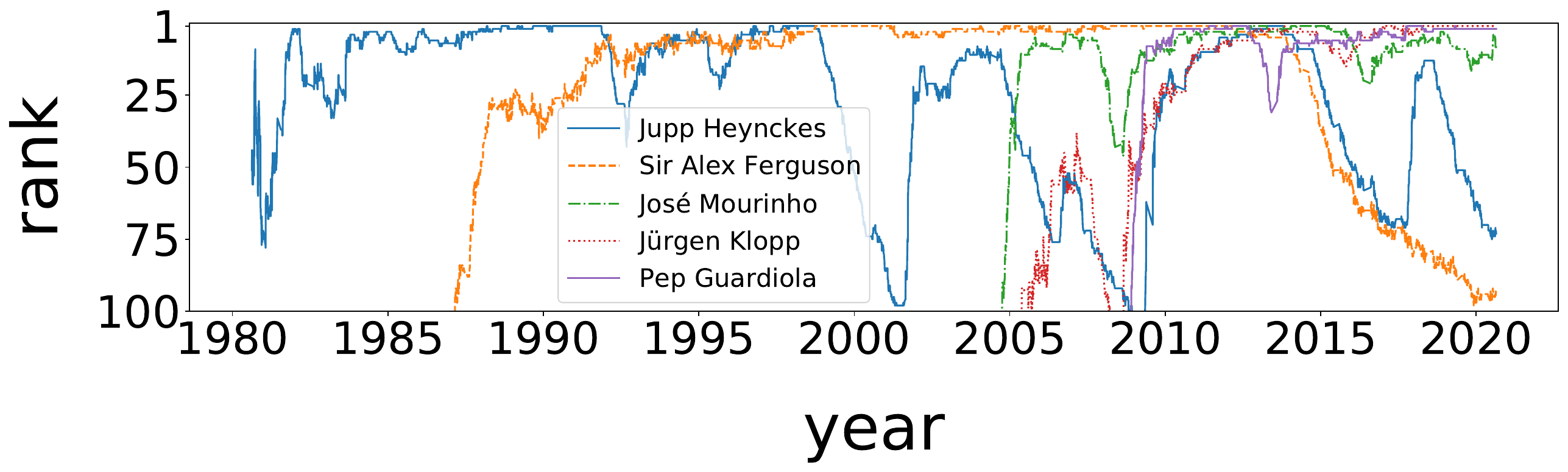}
\caption{
{\bf Comparing career performance of soccer coaches.} We display career trajectories for five selected coaches: Jupp Heynckes, Sir Alex Ferguson, José Mourinho, Jürgen Klopp, and Pep Guardiola. 
}
\label{fig:6}
\end{figure}

Dynamic scores can be used to compare the performance of coaches at any given point in time. In Figure~\ref{fig:6} for example, we display dynamic rank positions of five selected coaches. We see that Jupp Heynckes and Sir Alex Ferguson top the ranking for long periods of time. After retirement, rank positions are lost exponentially fast due to the choice we made for the kernel function of Eq.~(\ref{eq:ind_edge}). The most recent seasons are instead dominated by José Mourinho, Pep Guardiola, and Jürgen Klopp, who is ranked number one by the end of season $2019/20$.

%%%%%%%%%%%%%%%%%
%%%%%%%%%%%%%%%%%
%%%%%%%%%%%%%%%%%
%%%%%%%%%%%%%%%%%

\subsection{Basketball}

We repeat a similar analysis on the basketball dataset. First, we establish the all-time ranking by aggregating all games in our dataset. We consider two different sets of games: the union of ABA and NBA games, and NBA games only. In Eq.~(\ref{eq:ind_edge}), we use $\beta = 0$  and set $t$ equal to the day of the most recent game in the dataset. The list of the top $10$ coaches of all time is reported in Table~\ref{table:2_bball}. As already stressed for the all-time ranking of soccer coaches, also here we note that career longevity is strongly correlated with overall performance (see Figure~\ref{fig:a1}). The inclusion/exclusion of ABA games slightly modify the rank position of some coaches, although the names appearing in the top $10$ are basically the same irrespective of the particular dataset considered. Red Auerbach is elected as the best coach of all time.  He was the head coach of the Boston Celtics for more than $15$ seasons winning $9$ NBA titles. Auerbach is followed by Gregg Popovich, who is the current coach of the San Antonio Spurs. He has been coaching the same team for more than $20$ years. Under Gregg Popovich, except for his first and last seasons, the Spurs always made the playoffs, they never fell below $50\%$ win percentage, and won $5$ NBA championships. In the third place is Larry Brown, who had a long career both in the ABA and NBA. He coached in both leagues for more than $25$ seasons, winning $1$ NBA championship, and reaching $2$ NBA and $1$ ABA Finals. When we consider NBA games only and exclude ABA, Larry Brown drops from the third to the tenth place in the ranking. The third place in the all-time ranking based on NBA games only is occupied by Phil Jackson, head coach of the Chicago Bulls during the $1990$s and of the Los Angeles Lakers in two separate periods, winner of $11$ NBA titles.

\begin{table}[!htb]
\begin{center}
\begin{tabular}{|r|c|c|}\hline
%\begin{tabular}{|r|c|c|}\hline 
Rank & ABA + NBA & NBA only  \\\hline
1 & Red Auerbach & Red Auerbach \\\hline
2 & Gregg Popovich & Gregg Popovich \\\hline
3 & Larry Brown & Phil Jackson \\\hline
4 & Don Nelson & Don Nelson \\\hline
5 & Phil Jackson & Lenny Wilkens \\\hline
6 & Lenny Wilkens & Jerry Sloan \\\hline
7 & Jerry Sloan & Pat Riley \\\hline
8 & Pat Riley & George Karl \\\hline
9 & George Karl & John Kundla \\\hline
10& John Kundla & Larry Brown \\\hline
\end{tabular}
\end{center}
\caption{{\bf Top $10$ coaches in American basketball of all time.} 
We report the $10$ best coaches using ABA + NBA games, and then using NBA games only. The difference between the two networks considered are games played in the ABA between seasons $1967/68$ and $1975/76$ (see Table~\ref{table:1_bball}). In both cases, the games considered in the analysis were played from the $1946/47$ season till the $2019/20$ season.  Weighted networks of contact among coaches are constructed by setting $\beta =0$ in Eq.~(\ref{eq:ind_edge}).}
\label{table:2_bball}
\end{table}

%We also divided the data into decades for NBA, and ranked coaches for only the games in the respective decades. The results for the best performing coaches in NBA in each decade since 50s can be seen in Table \ref{table:4}. One of the notable observations is Phil Jackson having the best performance in 90s. He was the coach of 6-time NBA champions Chicago Bulls, and all those championships were achieved during the 90s. Another thing that can be pointed out should be the dominance of Gregg Popovich, who was ranked the best coach in the last two decades. 

The top coaches of the decade are: Red Auerbach in the $1950$s, Alex Hannum in the $1960$s, Dick Motta in the $1970$s, Pat Riley in the $1980$s, Phil Jackson in the $1990$s, and Gregg Popovich in the $2000$s and $2010$s.

The top coaches of the season are reported in Table~\ref{table:4_bball}. We see  that coaches making in the all-time top $10$ ranking are topping the ranking in several seasons. Generally, the top coach of the season corresponds to the coach of the NBA champion team. The one-to-one map is more apparent in basketball than in soccer due to the structure of the NBA basketball tournament. Top-performing basketball teams play a high number of post-season games; in national soccer championships, the number of games is the same for all teams.

We finally take advantage of dynamic weights to establish the list of the top $10$ coaches currently managing NBA teams. Erik Spoelstra is at the top of the list. As the coach of Miami Heat, he reached the NBA Finals in the 2019/20 season. The rest of the ranking is: Doc Rivers, Brad Stevens, Mike D'Antoni, Mike Malone, Frank Vogel, Gregg Popovich, Mike Budenholzer, Nick Nurse, and Terry Stotts. 

\begin{figure}[!htb]
\centering
\includegraphics[width=.45\textwidth]{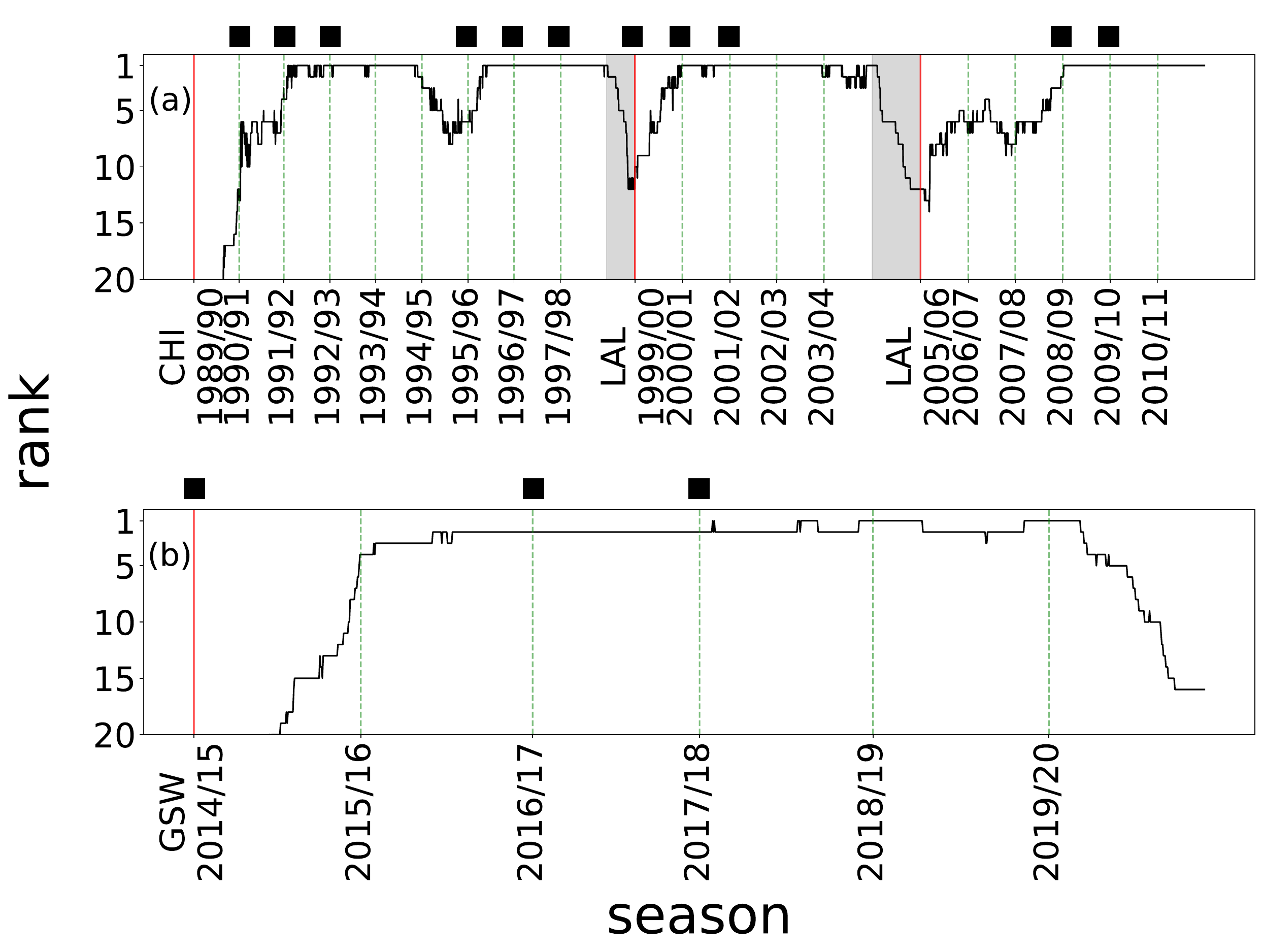}
\caption{
{\bf Monitoring the career performance of basketball coaches.} (a) 
We visualize the dynamic rank of Phil Jackson throughout his career. Rank positions are updated after each game day. Red lines indicate when a coach takes over a new team, dashed green lines indicate the start of a new season with his current team, and grey shaded areas represent periods of time when the coach is not managing any team. A black square indicates that the coach won the NBA championship in the corresponding season.
CHI stands for Chicago Bulls, and LAL for Los Angeles Lakers. (b) Same as in panel a, but for Steve Kerr. GSW stands for Golden State Warriors.}
\label{fig:5_bball}
\end{figure}

Dynamic weighted networks are further used to monitor the career evolution of coaches, as done for example in Figure~\ref{fig:5_bball}, where we display the career trajectories of Phil Jackson and Steve Kerr. 

Phil Jackson started his career at the Chicago Bulls, where he won 6 NBA titles. Later, he won 5 more NBA titles with the Los Angeles Lakers in two separate periods. Throughout his career, he  is consistently ranked in the top 5 (very often at the very top), according to his dynamic score. The only period when he drops out of the top 5 ranking, except for the periods when he was not coaching, is around 1994/95, the second season of the first retirement of Michael Jordan from basketball.

The career trajectory of Steve Kerr, although short, is very successful. In his 6 seasons with Golden State Warriors, he reached 5 NBA Finals and won 3 NBA titles. According to dynamic CoachScore, he enters in the top 5 ranking at the start of his second season and does not drop out of the top 5 ranking until his last season. 

\begin{figure}[!htb]
\centering
\includegraphics[width=.45\textwidth]{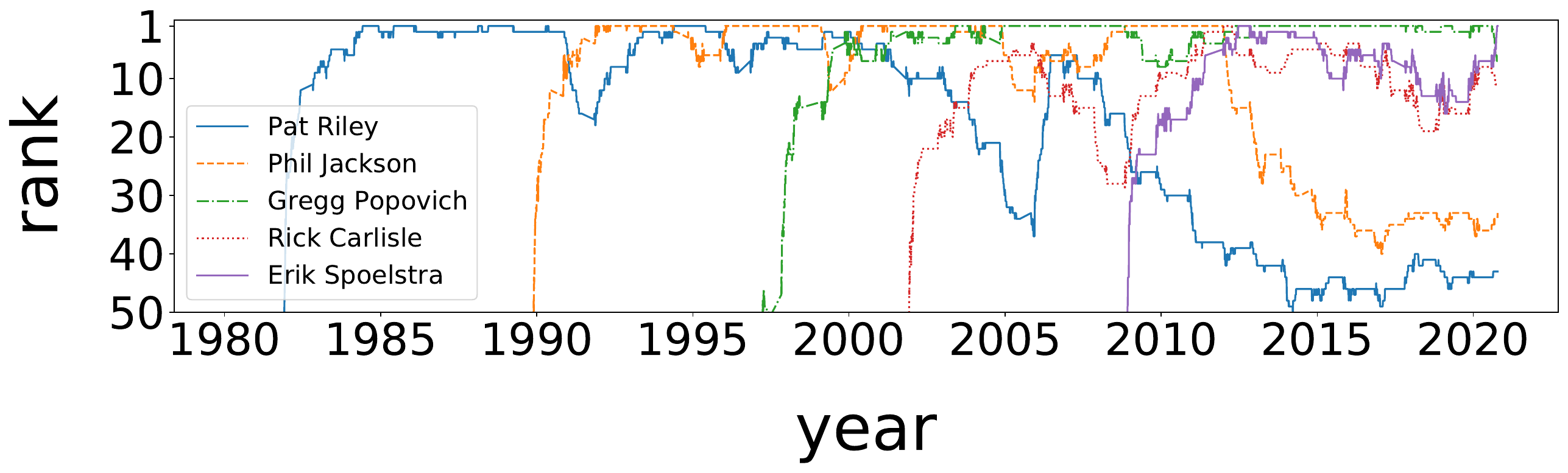}
\caption{{\bf Comparing career performance of basketball coaches.} We display career trajectories for five selected coaches: Pat Riley, Phil Jackson, Gregg Popovich, Rick Carlisle, and Erik Spoelstra.}
\label{fig:6_bball}
\end{figure}

Dynamic rank is also used to compare the performance of different coaches at the same instant of time, as done for example in Figure~\ref{fig:6_bball}. Here, we see Pat Riley at the top of the ranking around the mid $1980$s and the mid $1990$s. Phil Jackson and Gregg Popovich both have long runs at the top of the ranking for around $20$ years. Rick Carlisle and Erik Spoelstra reach the top $5$ ranking in the $2010$s. Rick Carlisle won a NBA title in 2010/11 with Dallas Mavericks. Erik Spoelstra won 2 titles in 2011/12 and 2012/13 with Miami Heat. %, where the squad included LeBron James. 

%%%%%%%%%%%%%%%%%%%%%%%%%%%%%%%%%%%%%%%%%%%%%%%%%%%%%%%%%%%%%%%%%%%%%%%%%%%%%%%%%%%%%%%%%%%%%%%%%%%%%%%%%%%%%%%%%%%%%%%%%%%%%%%%%%%%%%%%%%%%%%%%%%%%%%%%%%%%%%%%%%%%%%%%%%%%%%%%%%%%%%%%%%%%%%%%%%%%%%%%%%%%%%%%%%%%%%%%%%%%%

\section{Conclusions}
\label{conclusions}

The proper evaluation of the career of a professional soccer coach should account for a myriad of factors, e.g., the strength of the teams coached and the level of difficulty of the competitions where the coach participates in. However, these factors are hardly measurable, making the task of gauging performance on the basis of trophies or other achievements very challenging. For instance, should one value more a title in the Italian Serie A in the early $2000$s, in the Spanish La Liga in the mid $2010$s, or in the English Premier League today?  Also, the challenge is exacerbated by the fact that a career may span tens of seasons, and involve multiple teams and leagues. Cases like Sir Alex Ferguson, who managed the same team for $20+$ years, are quite rare. Even for these special cases, the high variability of the rest of the system makes extremely difficult to account for all the potential factors that one should quantify when assessing their career performance.

As only one league exists, gauging the performance of coaches of professional American basketball teams seems easier than it is for coaches of European soccer clubs. However, also in American basketball, factors that are important to measure the career performance of a coach vary on a time scale much shorter than the duration of the coach's career. Thus, attempts to compare coaches on the basis of simple counting strategies -- e.g., number of wins, number of trophies -- may not be completely fair, as the value of individual events may not be comparable from season to season, especially over extended periods of time.

In this paper, we avoid to explicitly give values to specific events. We just let the system decide the importance of the events in a self-consistent manner. Our approach is based on a macroscopic perspective of sports competitions. A game between two teams is seen as an elementary interaction among their respective coaches, with the direction of the interaction depending on the game outcome. The aggregation of data from many games allows us for the construction of a web of contacts among coaches. We use PageRank centrality, here renamed as CoachScore, to self-consistently determine the relative performance of a coach in the system. 
%\ch{The flexibility of our approach allows us to be able to consider different time frames and competitions. On the other hand, the number of losses do not have a significant affect on the score of a coach. This can be overcome using different measures \cite{masuda2009impact}, but it would also be measuring another dimension of the performance. Overall, the choice of the method depends on the dimension that we want to measure.}
%The only important ingredient of our ranking recipe is given by the set of games used to construct the network. %We consider different possibilities, aggregating data of multiple seasons and/or leagues. 

We do not claim that our way of quantifying  performance is better than others. 
For instance, we are aware of the limitations in the use of PageRank in dense networks~\cite{medo20}. Also, PageRank displays a strong age dependence when used in growing/dynamic networks~\cite{medo15, medo17}, and has the tendency of weighing losses much less than wins~\cite{masuda2009impact}. Different metrics of performance can be used to alleviate the above issues. However, we expect that any metric is affected by some limitations that narrow its usage and allow to properly gauge some specific features of performance only. We indeed believe that no single metric should be used to make direct comparisons among coaches, as performance is a multidimensional object. Our proposed score can be seen as one of these dimensions. With these considerations in mind, we hope that sports fans could enjoy the additional results provided in the companion website \web~ that is integral part of the present work. 

\section*{Funding}
This work was partially supported by National Science Foundation [CMMI-1552487].

\section*{Acknowledgements}
The authors thank D. Mazzilli for comments on the manuscript. 
%The authors acknowledge partial support from the National Science Foundation (CMMI-1552487). 

%%%%%%%%%%%%%%%%%%%%%%%%%%%%%%%%%%%%%%%%%%%%%%%%%%%%%%%%%%%%%%%%%%%%%%%%%%%%%%%%%%%%%%%%%%%%%%%%%%%%%%%%%%%%%%%%%%%%%%%%%%%%%%%%%%%%%%%%%%%%%%%%%%%%%%%%%%%%%%%%%%%%%%%%%%%%%%%%%%%%%%%%%%%%%%%%%%%%%%%%%%%%%%%%%%%%%%%%%%%%%

%%%%%
\appendix

\section{List of abbreviations}

In Table~\ref{table:abbreviations}, we report full forms of abbreviations used in the paper.

\begin{table}[!htb]
\begin{center}
\begin{tabular}{|r|l|}\hline 
Abbreviation & Full form (or translation) \\\hline
UEFA & Union of European Football Associations
\\\hline
FC / CF & Football Club \\\hline
AJ & Youth Association
\\\hline
AS & Sport Association% Association Sportive 
\\\hline
NBA & National Basketball Association \\\hline
ABA & American Basketball Association \\\hline
h2h & head-to-head\\\hline
\end{tabular}
\end{center}
\caption{{\bf List of abbreviations used in the paper.} From left to right, we report the abbreviation and the corresponding full form.}
\label{table:abbreviations}
\end{table}

\section{Additional results}
\label{appendix}
\begin{figure}[!htb]
\centering
\includegraphics[width=0.45\textwidth]{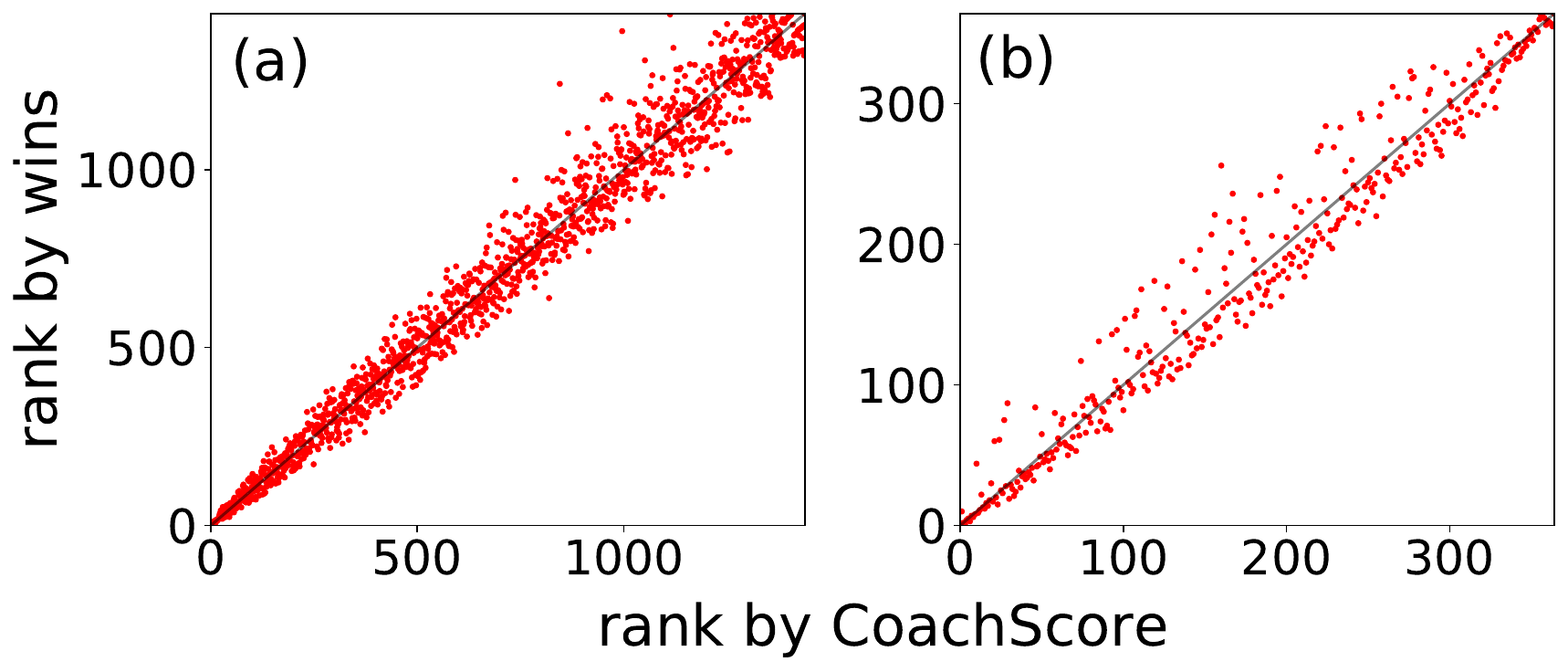}
\caption{{\bf Comparison of rankings by CoachScore and number of wins.} (a) We report the rank position of all soccer coaches in our combined dataset (see Table~\ref{table:1}). Rankings are performed by either relying on the CoachScore values or the number of wins. Each point in the plot is a coach. Spearman correlation coefficient is $\rho=0.989$, while Kendall correlation coefficient is $\tau=0.923$. (b) Similar to panel a, but for basketball coaches. Rank correlation coefficients are $\rho=0.979$ and $\tau=0.888$.}
\label{fig:a1}
\end{figure}

\begin{figure}[!htb]
\centering
\includegraphics[width=0.45\textwidth]{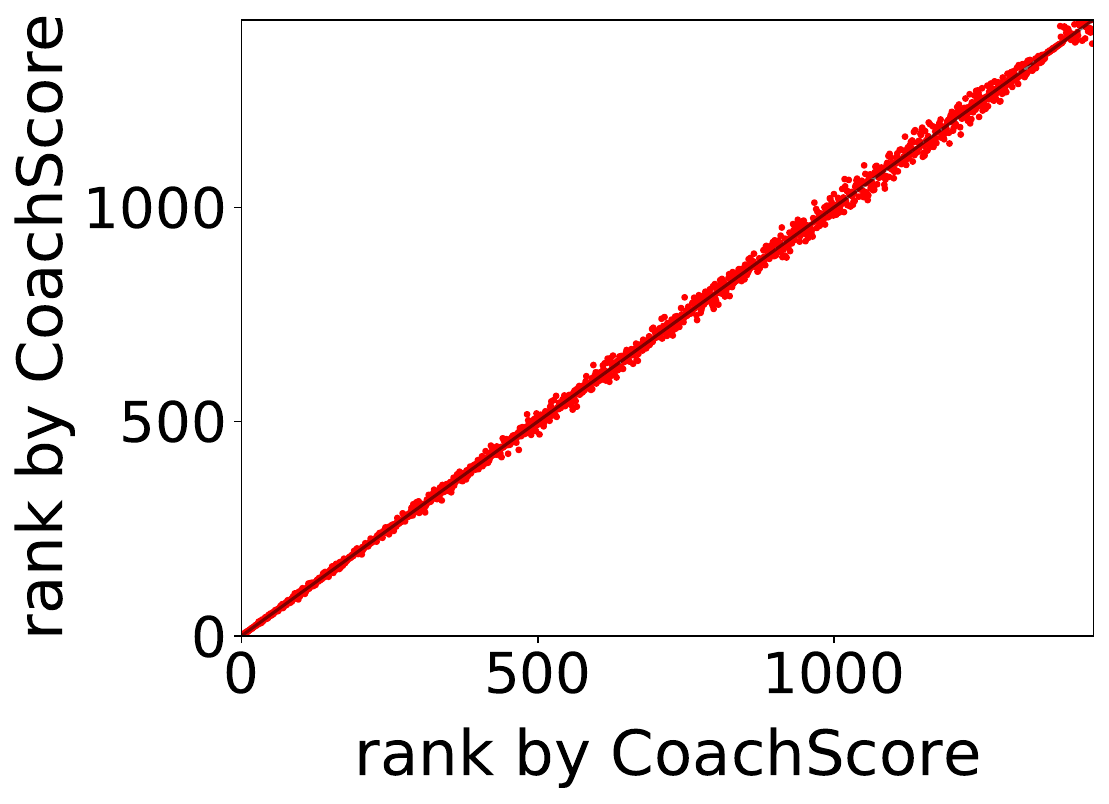}
\caption{{\bf Comparison of rankings by CoachScore obtained by using different weights for a loss in soccer.} We report the rank position of all soccer coaches of our combined dataset. The ranking on the x-axis relies on the network where a win has twice the weight of a tie 
[i.e., $q_{\textrm{tie}}=1$ and $q_{\textrm{loss}}=2$ in Eq.~(\ref{eq:ind_edge})]. The ranking on the y-axis relies on a network where a win has three times the weight of a tie
[i.e., $q_{\textrm{tie}}=1$ and $q_{\textrm{loss}}=3$ in Eq.~(\ref{eq:ind_edge})]. 
Spearman correlation coefficient is $\rho=0.999$, while Kendall correlation coefficient is $\tau=0.983$.}
\label{fig:a2}
\end{figure}

\begin{figure}[!htb]
\centering
\includegraphics[width=0.45\textwidth]{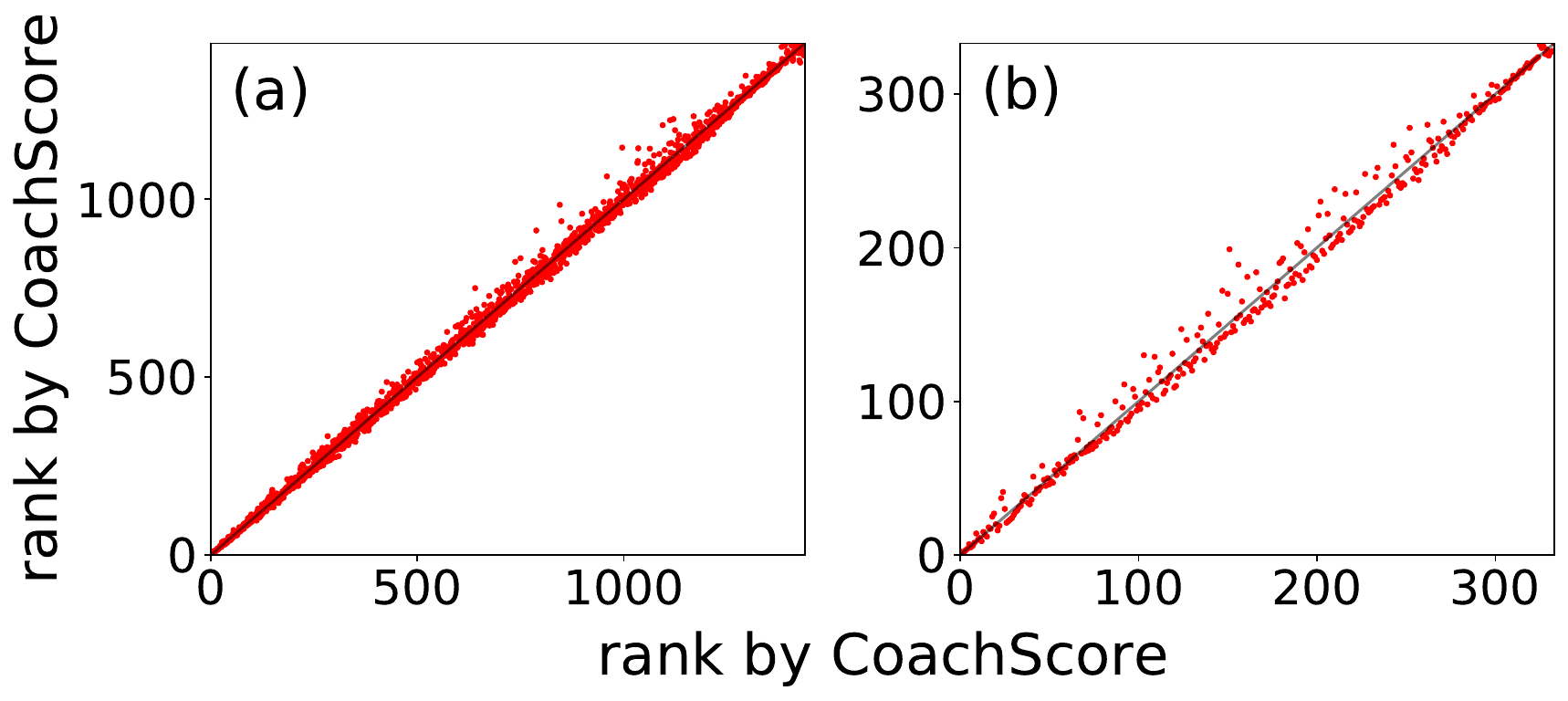}
\caption{{\bf Comparison of rankings by CoachScore while using different damping factors.} (a) We report the rank position of all soccer coaches of our combined dataset. The ranking on the x-axis is for CoachScore with damping factor $\alpha=0.85$, and the ranking on the y-axis with $\alpha=0.95$. Spearman correlation coefficient is $\rho=0.999$, while Kendall correlation coefficient is $\tau=0.973$. (b) Same as in panel a, but basketball coaches in NBA. Spearman correlation coefficient is $\rho=0.996$, while Kendall correlation coefficient is $\tau=0.954$.}
\label{fig:a3}
\end{figure}

In Figures~\ref{fig:a1}, ~\ref{fig:a2}, and ~\ref{fig:a3}, we report on additional results from our analysis. Specifically, we consider rank correlation plots between:  CoachScore and number of wins (Fig.~\ref{fig:a1}); CoachScore computed on networks with different ratios of win/tie (Fig.~\ref{fig:a2}); CoachScore computed for different values of the parameter $\alpha$ (Fig.~\ref{fig:a3}). In Tables \ref{table:4}-\ref{table:4a}, we report the top coaches for all seasons covered in our dataset in soccer. In Table \ref{table:4_bball}, we report the top coaches for all seasons in NBA.

%In Figure \ref{fig:a1}, we report both the rankings of coaches created by CoachScore and the number of wins. We observe a high correlation between the two rankings as expected. However, there are many differences between the two rankings, indicating that CoachScore is not the same as simply counting the number of wins of a coach.

\clearpage

%In Figure \ref{fig:a2}, we compare the coach rank positions in our combined dataset of soccer, when different weights are used for a win. Even though there are some differences between the two rankings, they show a high correlation, indicating any acceptable choice for the weight will yield similar results.

%In Figure \ref{fig:a3}, we compare the rank positions of coaches in both soccer and basketball when using different damping factors $\alpha$ for CoachScore. While there are changes in the ranking, the rank correlations remain high. One important change is in NBA, where Gregg Popovich passes Red Auerbach and tops the ranking if we use $\alpha=0.95$.

\begin{table*}[!htb]
\begin{center}
\begin{tabular}{|l|l|l|l|l|l|l|}\hline 
%\begin{tabular}{|l|l|l|l|l|l|l|}\hline 
Season & England & France & Germany & Italy & Spain & Combined \\\hline
1960/61 & - & - & - & Paolo Todeschini & Miguel Munoz & - \\\hline
1961/62 & - & - & - & Nereo Rocco & Miguel Munoz & - \\\hline
1962/63 & - & - & - & Helenio Herrera & Miguel Munoz & - \\\hline
1963/64 & - & - & Georg Knöpfle & Helenio Herrera & Miguel Munoz & - \\\hline
1964/65 & - & - & Willi Multhaup & Helenio Herrera & Roque Olsen & - \\\hline
1965/66 & - & - & Max Merkel & Bruno Pesaola & Roque Olsen & - \\\hline
1966/67 & - & - & Helmuth Johannsen & Heriberto Herrera & Jenő Kalmár & - \\\hline
1967/68 & - & - & Max Merkel & Nereo Rocco & Salvador Artigas & - \\\hline
1968/69 & - & - & Branko Zebec & Bruno Pesaola & Miguel Munoz & - \\\hline
1969/70 & - & - & Hennes Weisweiler & Manlio Scopigno & Marcel Domingo & - \\\hline
1970/71 & Don Revie & - & Hennes Weisweiler & Nereo Rocco & Alfredo di Stéfano & - \\\hline
1971/72 & Don Revie & - & Udo Lattek & Cestmir Vycpalek & Alfredo di Stéfano & - \\\hline
1972/73 & Bertie Mee & - & Udo Lattek & Cestmir Vycpalek & José Santamaría & - \\\hline
1973/74 & Don Revie & - & Udo Lattek & Cestmir Vycpalek & Rinus Michels & - \\\hline
1974/75 & Dave Mackay & - & Kuno Klötzer & Carlo Parola & Miljan Miljanic & - \\\hline
1975/76 & Dave Sexton & - & Udo Lattek & Luigi Radice & Miljan Miljanic & - \\\hline
1976/77 & Bob Paisley & - & Friedel Rausch & Luigi Radice & Luis Aragonés & - \\\hline
1977/78 & Brian Clough & - & Hennes Weisweiler & Giovanni Trapattoni & Luis Molowny & - \\\hline
1978/79 & Bob Paisley & - & Jürgen Sundermann & Nils Liedholm & Luis Molowny & - \\\hline
1979/80 & Bob Paisley & - & Branko Zebec & Eugenio Bersellini & Alberto Ormaetxea & - \\\hline
1980/81 & Ron Atkinson & Jean Vincent & Pál Csernai & Giovanni Trapattoni & Vujadin Boskov & Sir Bobby Robson \\\hline
1981/82 & Sir Bobby Robson & Gérard Banide & Ernst Happel & Giancarlo De Sisti & Alberto Ormaetxea & Udo Lattek \\\hline
1982/83 & Bob Paisley & Jean-Claude Suaudeau & Otto Rehhagel & Nils Liedholm & Alfredo di Stéfano & Ernst Happel \\\hline
1983/84 & Joe Fagan & Aimé Jacquet & Jupp Heynckes & Giovanni Trapattoni & Alfredo di Stéfano & Otto Rehhagel \\\hline
1984/85 & Howard Kendall & Aimé Jacquet & Udo Lattek & Osvaldo Bagnoli & Terry Venables & Aimé Jacquet \\\hline
1985/86 & Sir Kenny Dalglish & Gérard Houllier & Erich Ribbeck & Sven-Göran Eriksson & Luis Molowny & Luis Molowny \\\hline
1986/87 & Howard Kendall & Aimé Jacquet & Udo Lattek & Ottavio Bianchi & Terry Venables & Leo Beenhakker \\\hline
1987/88 & Sir Kenny Dalglish & Arsène Wenger & Otto Rehhagel & Arrigo Sacchi & Leo Beenhakker & Leo Beenhakker \\\hline
1988/89 & George Graham & Arsène Wenger & Jupp Heynckes & Giovanni Trapattoni & Leo Beenhakker & Ottavio Bianchi \\\hline
1989/90 & Sir Kenny Dalglish & Gérard Gili & Jupp Heynckes & Arrigo Sacchi & John Toshack & John Toshack \\\hline
1990/91 & George Graham & Arsène Wenger & Jupp Heynckes & Vujadin Boskov & Javier Irureta & Karl-Heinz Feldkamp \\\hline
1991/92 & Howard Wilkinson & Arsène Wenger & Dragoslav Stepanovic & Fabio Capello & Luis Aragonés & Christoph Daum \\\hline
1992/93 & Sir Alex Ferguson & Arsène Wenger & Otto Rehhagel & Fabio Capello & Johan Cruyff & Johan Cruyff \\\hline
1993/94 & Sir Alex Ferguson & Artur Jorge & Friedel Rausch & Fabio Capello & Arsenio Iglesias & Fabio Capello \\\hline
1994/95 & Sir Kenny Dalglish & Jean-Claude Suaudeau & Otto Rehhagel & Marcello Lippi & Jorge Valdano & Marcello Lippi \\\hline
1995/96 & Sir Alex Ferguson & Patrice Bergues & Ottmar Hitzfeld & Fabio Capello & Radomir Antić & Ottmar Hitzfeld \\\hline
1996/97 & Sir Alex Ferguson & Jean Tigana & Giovanni Trapattoni & Marcello Lippi & Fabio Capello & Sir Bobby Robson \\\hline
1997/98 & Arsène Wenger & Daniel Leclercq & Otto Rehhagel & Marcello Lippi & Bernd Krauss & Radomir Antić \\\hline
1998/99 & Arsène Wenger & Élie Baup & Ottmar Hitzfeld & Alberto Zaccheroni & Louis van Gaal & Ottmar Hitzfeld \\\hline
1999/00 & Sir Alex Ferguson & Claude Puel & Christoph Daum & Sven-Göran Eriksson & Txetxu Rojo & Sven-Göran Eriksson \\\hline
2000/01 & Sir Alex Ferguson & Jacques Santini & Huub Stevens & Fabio Capello & Luis Aragonés & Ottmar Hitzfeld \\\hline
2001/02 & Arsène Wenger & Jacques Santini & Ottmar Hitzfeld & Fabio Capello & Rafael Benítez & Javier Irureta \\\hline
2002/03 & Sir Alex Ferguson & Paul Le Guen & Ottmar Hitzfeld & Marcello Lippi & Vicente del Bosque & Vicente del Bosque \\\hline
2003/04 & Arsène Wenger & Vahid Halilhodzic & Thomas Schaaf & Carlo Ancelotti & Rafael Benítez & Javier Irureta \\\hline
2004/05 & José Mourinho & Paul Le Guen & Falko Götz & Roberto Mancini & Frank Rijkaard & Carlo Ancelotti \\\hline
2005/06 & Sir Alex Ferguson & Gérard Houllier & Thomas Doll & Fabio Capello & Frank Rijkaard & Frank Rijkaard \\\hline
2006/07 & Sir Alex Ferguson & Gérard Houllier & Armin Veh & Roberto Mancini & Fabio Capello & Roberto Mancini \\\hline
2007/08 & Sir Alex Ferguson & Laurent Blanc & Ottmar Hitzfeld & Roberto Mancini & Bernd Schuster & Sir Alex Ferguson \\\hline
2008/09 & Rafael Benítez & Eric Gerets & Ralf Rangnick & José Mourinho & Pep Guardiola & Pep Guardiola \\\hline
\end{tabular}
\end{center}
\caption{
%{\bf Best coaches in soccer per season based on PageRank.} 
{\bf Top European soccer coaches of the season.} 
For each season, we report the best coach of each of the national leagues we consider in this paper (see Table~\ref{table:1}). Empty cells indicate that no data are at our disposal for the specific combination of league/season. In the rightmost column, we report the best coach of each season obtained on the basis of the combination of all data at our disposal, including national and international competitions.}
\label{table:4}
\end{table*}

\begin{table*}[!htb]
\begin{center}
\begin{tabular}{|l|l|l|l|l|l|l|}\hline 
%\begin{tabular}{|l|l|l|l|l|l|l|}\hline 
Season & England & France & Germany & Italy & Spain & Combined \\\hline
2009/10 & Carlo Ancelotti & Jean Fernandez & Louis van Gaal & Claudio Ranieri & Pep Guardiola & José Mourinho \\\hline
2010/11 & Sir Alex Ferguson & Rudi Garcia & Jürgen Klopp & Massimiliano Allegri & Pep Guardiola & Pep Guardiola \\\hline
2011/12 & Roberto Mancini & René Girard & Jürgen Klopp & Antonio Conte & José Mourinho & Pep Guardiola \\\hline
2012/13 & Sir Alex Ferguson & Carlo Ancelotti & Jürgen Klopp & Antonio Conte & José Mourinho & Jupp Heynckes \\\hline
2013/14 & José Mourinho & Laurent Blanc & Jürgen Klopp & Antonio Conte & Diego Simeone & Diego Simeone \\\hline
2014/15 & José Mourinho & Laurent Blanc & Dieter Hecking & Massimiliano Allegri & Luis Enrique & Luis Enrique \\\hline
2015/16 & Claudio Ranieri & Laurent Blanc & Pep Guardiola & Massimiliano Allegri & Diego Simeone & Luis Enrique \\\hline
2016/17 & Jürgen Klopp & Leonardo Jardim & Carlo Ancelotti & Massimiliano Allegri & Luis Enrique & Zinédine Zidane \\\hline
2017/18 & José Mourinho & Unai Emery & Ralph Hasenhüttl & Massimiliano Allegri & Ernesto Valverde & Zinédine Zidane \\\hline
2018/19 & Pep Guardiola & Thomas Tuchel & Lucien Favre & Massimiliano Allegri & Ernesto Valverde & Ernesto Valverde \\\hline
2019/20 & Jürgen Klopp & Thomas Tuchel & Julian Nagelsmann & Maurizio Sarri & Zinédine Zidane & Julian Nagelsmann \\\hline
\end{tabular}
\end{center}
\caption{
%{\bf Best coaches in soccer per season based on PageRank.} 
{\bf Top European soccer coaches of the season.} Continuation of Table~\ref{table:4}.}
\label{table:4a}
\end{table*}

\begin{table*}[!htb]
\begin{center}
\begin{tabular}{|c|l||c|l|}\hline 
%\begin{tabular}{|r|l||r|l|}\hline 
Season & NBA & Season & NBA \\\hline
$1946/47$ & Red Auerbach & $1983/84$ & K.C. Jones \\\hline
$1947/48$ & Buddy Jeannette & $1984/85$ & K.C. Jones \\\hline
$1948/49$ & John Kundla & $1985/86$ & K.C. Jones \\\hline
$1949/50$ & John Kundla & $1986/87$ & K.C. Jones \\\hline
$1950/51$ & Les Harrison & $1987/88$ & Pat Riley \\\hline
$1951/52$ & John Kundla & $1988/89$ & Chuck Daly \\\hline
$1952/53$ & John Kundla & $1989/90$ & Rick Adelman \\\hline
$1953/54$ & John Kundla & $1990/91$ & Mike Dunleavy \\\hline
$1954/55$ & Al Cervi & $1991/92$ & Phil Jackson \\\hline
$1955/56$ & George Senesky & $1992/93$ & Phil Jackson \\\hline
$1956/57$ & Red Auerbach & $1993/94$ & Rudy Tomjanovich \\\hline
$1957/58$ & Red Auerbach & $1994/95$ & Rudy Tomjanovich \\\hline
$1958/59$ & Red Auerbach & $1995/96$ & Phil Jackson \\\hline
$1959/60$ & Red Auerbach & $1996/97$ & Phil Jackson \\\hline
$1960/61$ & Red Auerbach & $1997/98$ & Phil Jackson \\\hline
$1961/62$ & Red Auerbach & $1998/99$ & Gregg Popovich \\\hline
$1962/63$ & Red Auerbach & $1999/00$ & Phil Jackson \\\hline
$1963/64$ & Red Auerbach & $2000/01$ & Phil Jackson \\\hline
$1964/65$ & Red Auerbach & $2001/02$ & Phil Jackson \\\hline
$1965/66$ & Red Auerbach & $2002/03$ & Gregg Popovich \\\hline
$1966/67$ & Alex Hannum & $2003/04$ & Phil Jackson \\\hline
$1967/68$ & Bill Russell & $2004/05$ & Gregg Popovich \\\hline
$1968/69$ & Bill Russell & $2005/06$ & Avery Johnson \\\hline
$1969/70$ & Red Holzman & $2006/07$ & Gregg Popovich \\\hline
$1970/71$ & Larry Costello & $2007/08$ & Doc Rivers \\\hline
$1971/72$ & Bill Sharman & $2008/09$ & Phil Jackson \\\hline
$1972/73$ & Red Holzman & $2009/10$ & Phil Jackson \\\hline
$1973/74$ & Larry Costello & $2010/11$ & Rick Carlisle \\\hline
$1974/75$ & K.C. Jones & $2011/12$ & Erik Spoelstra \\\hline
$1975/76$ & Tom Heinsohn & $2012/13$ & Gregg Popovich \\\hline
$1976/77$ & Jack Ramsay & $2013/14$ & Gregg Popovich \\\hline
$1977/78$ & Dick Motta & $2014/15$ & Steve Kerr \\\hline
$1978/79$ & Lenny Wilkens & $2015/16$ & Steve Kerr \\\hline
$1979/80$ & Billy Cunningham & $2016/17$ & Steve Kerr \\\hline
$1980/81$ & Bill Fitch & $2017/18$ & Steve Kerr \\\hline
$1981/82$ & Billy Cunningham & $2018/19$ & Nick Nurse \\\hline
$1982/83$ & Billy Cunningham & $2019/20$ & Frank Vogel \\\hline
\end{tabular}
\end{center}
\caption{
{\bf Top NBA coaches of the season.} 
We consider NBA games only. We report the name of the top
coach of the season according to our ranking. Weights of the network connections are obtained by setting $\beta = 0$ in Eq.~(\ref{eq:ind_edge}).}
\label{table:4_bball}
\end{table*}

\clearpage
\newpage

%%%%%

\end{document}